# Analytical study on dynamic responses of sheet-pile groin subjected to transient lateral impulse

**Tao Wu[a], Hong-lei Sun[b,*], Yuan-qiang Cai[a], Jun-tao Wu[a], Yun-peng Zhang[c]**

[a] Research Center of Coastal and Urban Geotechnical Engineering, College of Civil Engineering and Architecture, Zhejiang University, Hangzhou, 310058, PR China

[b] Institute of Geotechnical Engineering, College of Civil Engineering, Zhejiang University of Technology, Hangzhou 31000, PR China

[c] Zhejiang Institute, China University of Geosciences, Hangzhou, 311305, China

**Abstract**

A novel structure of sheet-pile groin consists of row piles and tied sheet has been introduced to resist the impact of tidal bore in recent years. However, the dynamic responses of the structure under tidal bore have not been carefully studied before. This paper proposed an analytical solution for the dynamic responses of the sheet-pile groin subjected to a transient lateral excitation. In order to take the flexural and longitudinal vibrations into consideration, the piles and the sheet are modeled by the Timoshenko beam (TB) theory and a one-dimensional rod, respectively. The soil around the piles is simulated by the dynamic Winkler foundation model. The excitation is considered by a lateral concentrated point load acting on the front pile. On this basis, the governing equations are constructed in time domain and solved in frequency domain, while the time-domain responses are finally obtained using the discrete Inverse Fourier

Transform. The accuracy of the presented analytical solution is validated by comparing with the results obtained from the FEM simulation and a model test. The effects of properties of the sheet, the piles and the soil on the dynamic responses are investigated through a comprehensive parametric study.

## 1. Introduction

A tidal bore is a positive surge that usually forms by a flood tide in shallow estuaries with large tidal ranges [1]. It induces a strong turbulent mixing in the estuarine zone which causes great damage to structures such as breakwaters in the flow region and on the near-shore region [2,3]. In order to reduce these adverse effects, many infrastructures have been built to protect the river embankments. Groins have been widely built perpendicularly to shorelines and river banks for embankment protection and accretion promotion. However, bucking failures, including head flushing and foot hanging, are often found in traditional rock-fill groins [4]. To overcome these deficiencies, a novel sheet-pile groin , as demonstrated in Fig.1, composed of two rows of piles along with sheet connecting the pile tops was proposed in engineering practices [5]. Compared to traditional groins, this newly proposed groin can protect the river banks and sea walls well to avoid bucking failures, and provide a much longer servicing period with fewer maintenance fees [6].

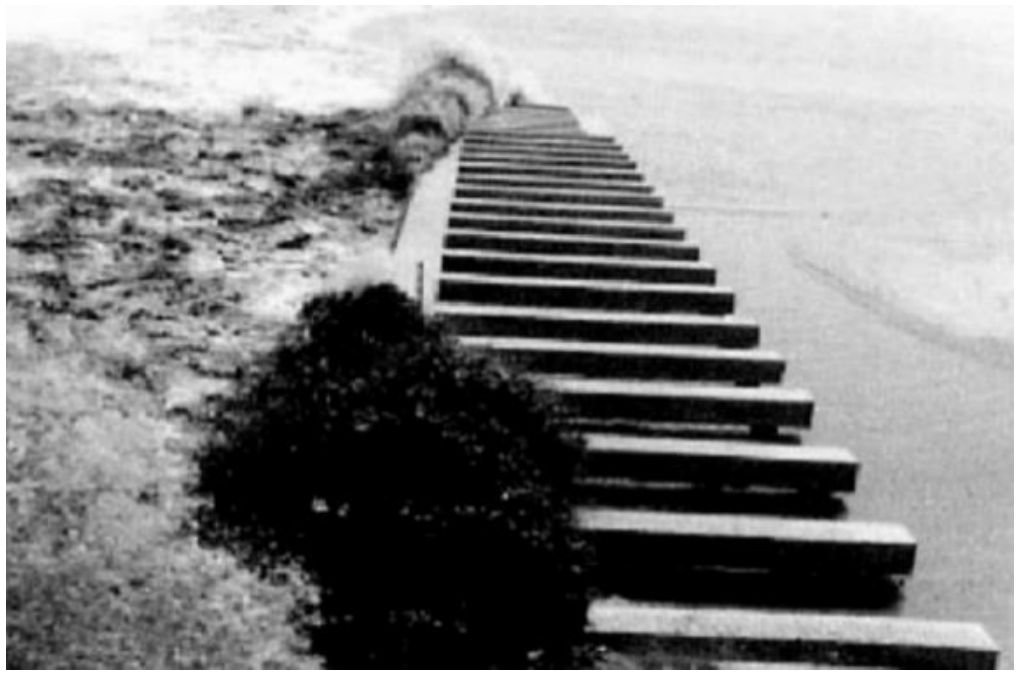

(a) Site map of the sheet-pile groin model

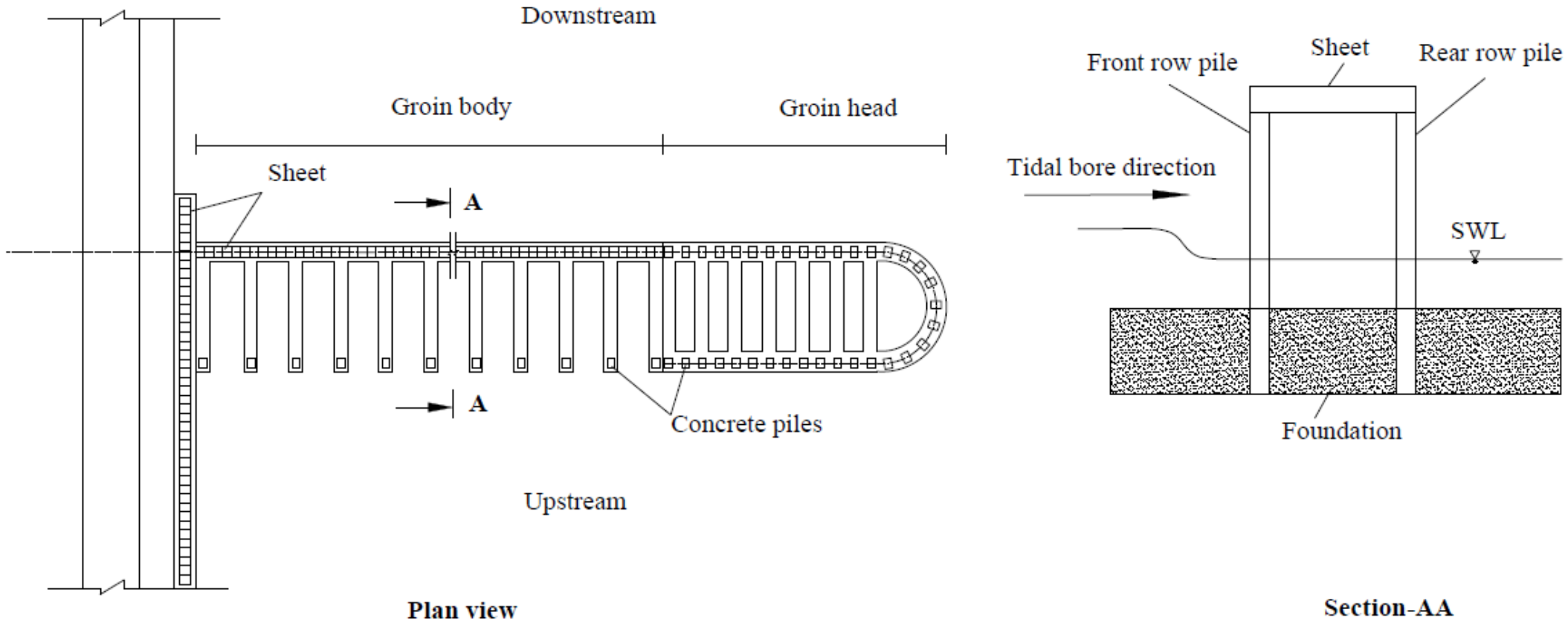

(b) Sketch of the sheet-pile groin model

Fig.1 sheet-pile groin

Many researchers have studied the effects of the tidal bore on hydraulic infrastructures. The tidal bore pressure and its influence factors were studied by Chen et al.[7] through field tests. Xu et al.[5] studied the dynamic interaction between the tidal bore and the sheet-pile groin structures, mainly focusing on the hydraulic characteristics of tide bore whereas the embedding of the pile is not considered. The dynamic pressure exerted by the tidal bore on the groin was measured by Vaghefi et al.[8] through model tests in a channel with a 90° bend. Mohapatra et al. [9] used a variant of the simplified marker and cell method to simulate the impact of a bore against a wall, and obtained the maximum force and the run-up height on the wall. Shen et al. [10] used a two-equation $k$-$\varepsilon$ model combined with a volume of fluid (VOF) method to simulate the propagation of cnoidal waves over a submerged bar. It can be concluded that previous studies mainly focused on the numerical simulations and model tests of the hydraulic characteristics of the sheet-pile groin. Moreover, the tidal bore causes vibrations of the sheet-pile groin structure, which will generate additional stress at the joints between the sheet and piles. Damages, such as the separations of the joints and the excessive horizontal displacement of the pile head [11–13], can be caused by these stresses , thus affecting the durability and increasing maintenance costs. However, the dynamic structural responses of the sheet-pile groin have not been well studied and elaborated.

There have existed various models among researchers' works for studying the dynamic responses of the piles[14–21]. Novak and Nogami [14,15] studied the lateral vibration of the piles in homogeneous and isotropic soil based on the continuous consideration of the pile-soil coupling interaction under the condition of small deformation. Dynamic Winkler model is widely utilized for soil dynamic reaction simulations for its simplicity and reliability [17–19]. Novak [17] and Gazetas [18] provided an analytical solution for the determination of the elastic and damping coefficients in dynamic Winkler model. The lateral dynamic responses of single pile or pile group can be derived owing to the above simplified soil models [19–23]. All above mentioned studies are mainly based on single pile, pile group and pile cap. The difference between piles with cap and sheet-pile groin is in three aspects. First, the horizontal deformation of the pile cap is small which introduces the hypothesis that all the displacements of pile heads are the same. Second, the joint between the cap and the pile head is rigid which limits the rotation of the pile head. Third, the mass of the cap is large which cannot be neglected in the analysis of the dynamic responses. However, in a sheet-pile groin due to the flexibility of the sheet, its axial vibration needs to be considered. The joint between the piles and the sheet is hinged which limits the bending moment but not the rotation of the pile head. The mass of the sheet is small which can be neglected in the analysis of the dynamic responses. Therefore, it is necessary to study the dynamic responses of the sheet-pile groin subjected to transient lateral impulse. Otherwise, the generalized model of the pile is derived from Euler-Bernoulli Beam (EB), which neglects the pile shear deformation. In most cases, the flat section assumption would perform well except for piles with small aspect ratio, which is exactly the most common piles applied in anti-flood groins [13]. Timoshenko further developed the beam theory by taking the shear deformation and rotary inertia into account [24]. Furthermore, in some groin-like structural studies, the sheet is simplified as a mass block, ignoring the compression displacement, axial force and the coupling action such as the transmission of the shear force and the moment between the front and the rear piles [25].

In this paper, the sheet-pile groin is analyzed with the piles modelled as a TB and the sheet as a rod. The interfaces between the piles and the sheet are governed by the continuous and compatibility deformation conditions. The dynamic Winkler model is adopted to simulate the soil around the pile. A transient lateral impulse is imposed at the front row pile and the dynamic responses of the groin is studied. The proposed analytical solution is validated through the comparisons with FEM. Finally, a detailed parametric study is given to exhibit the influence of various factors on the groin responses.

## 2. Analytical model and problem definition

The sheet-pile groin extends to the river channel along the length direction, and each pile is arranged equidistantly. The cross section of the row piles on the tide surface does not change along the length. The side surface is subjected to tidal bore loads parallel to the cross section and not changing along the length. It can be simplified as a plane problem that considered the sheet-pile groin as a single sheet-pile structure composed of row piles and sheet connected on the top. The detailed description of the simplified mathematical model is depicted in Fig.2. The row piles are partially buried in the soil and the sheet is connected to the top of piles. The system has no external force except the transient excitation acted on the front row pile. The $z$ axis is set at the center of the front row-pile while the positive direction is recognized to be upward. And the $x$ axis is set at the toe of the front row-pile while the positive direction points to the right, with the origin of the coordinate set at the intersection with the $z$ axis. The soil is simulated by homogeneous dynamic Winkler model. Two piles (front and rear row pile) with circular cross-section are divided into three segments along the length direction respectively. Pile segment 1 is embedded in the soil, whose lateral resistance is determined by the dynamic Winkler foundation model [19,20,22,23]. Pile segments 2 and 3 are above the ground, which are partitioned at the position of the same height extending along the point where the lateral excitation acts at. The top of the pile segment 3 and the end of the sheet are hinged. The lengths of pile segments are labelled as $h_1$, $h_2$, $h_3$ respectively. The intersections of the front row pile segments are marked as $A$, $B$, $C$ and $D$ respectively from the bottom to the top, whose corresponding coordination can be denoted according to the length conversion.

There existed several dynamic interactions between the pile and the sheet such as displacements, rotation angels, forces and moments, which satisfies the continuity and compatibility conditions. As many researchers have found in practical applications, the sheet is approximately translational and the row piles' rotation angels keeps constant [6,26–28]. Consequently in Fig.2, the shear force of the piles is considered to be equal to the axial force of the rod. And the displacement of the piles and sheet is continuous at their joints while ignoring the interactions of the rotation angels and the moments. It is supposed to use $f_q\left(t\right)$ and $f_u\left(t\right)$ to represent the shear force at the top of the front and rear row piles respectively. Also $f_q^{'}\left(t\right)$ and $f_u^{'}\left(t\right)$ are used to express the axial force at the left and right ends of the sheet.. The pile lateral vibration is accounted by employing the TB theory [25], which accounts for shear and rotational inertia. The sheet is considered as an axial viscoelastic rod. The deflection and dynamic equilibrium of the unit width of the infinitesimal pile element are depicted in Fig.3. In Fig.3, $M_{mn}$ and $Q_{mn}$ $(n=1, 2, 3)$ are the bending moment and the shearing force for pile segments 1–3, while $m=1, 2$ represents the front row pile and the rear row pile respectively; $\theta_{mn}(z,t)$ and $u_{mn}(z,t)$ denote the angle of rotation due to bending and the lateral deflection, respectively; $q$ denotes the soil lateral distributed stress acting on the piles.

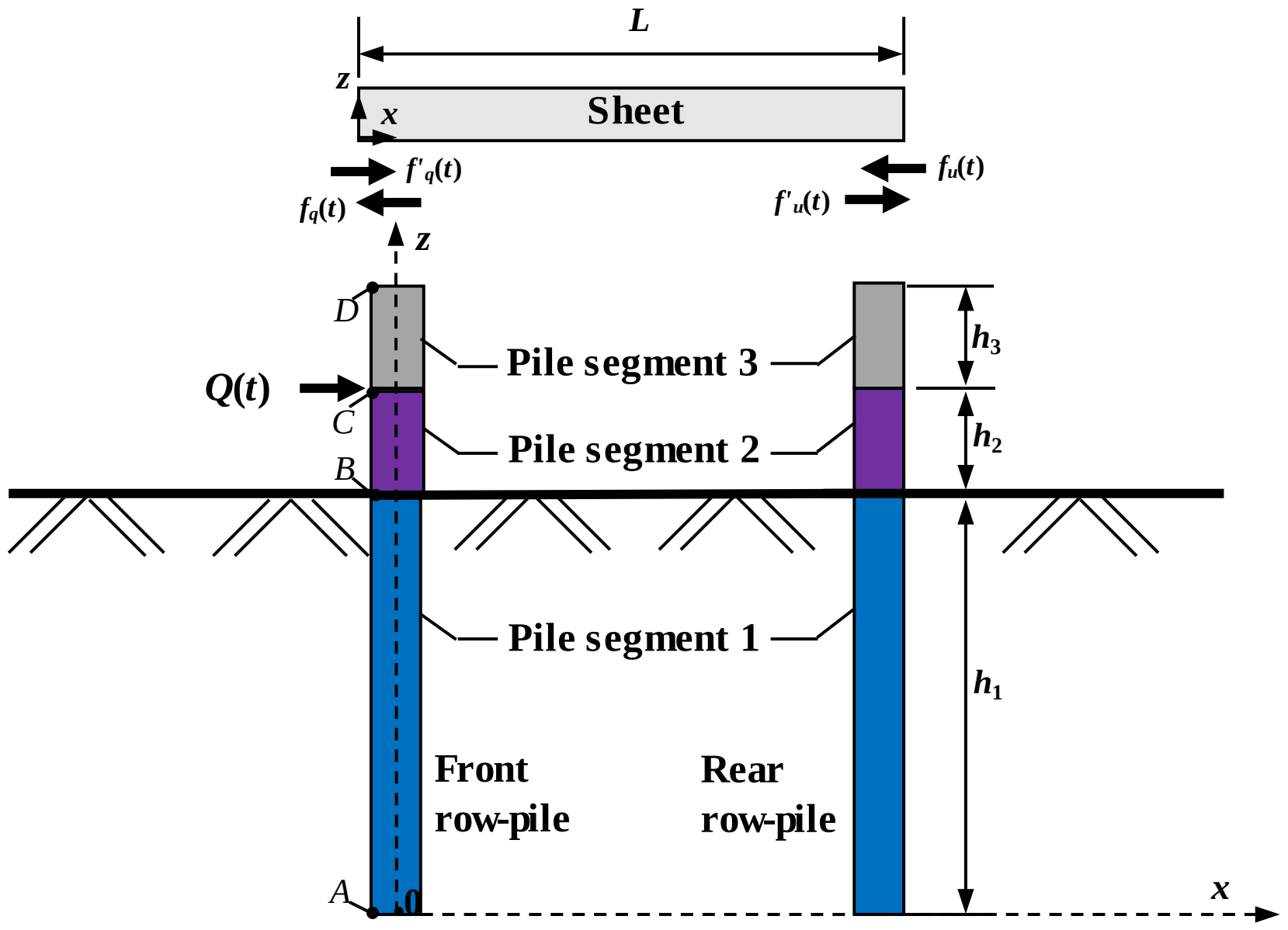

Fig.2 Schematic of analytical model

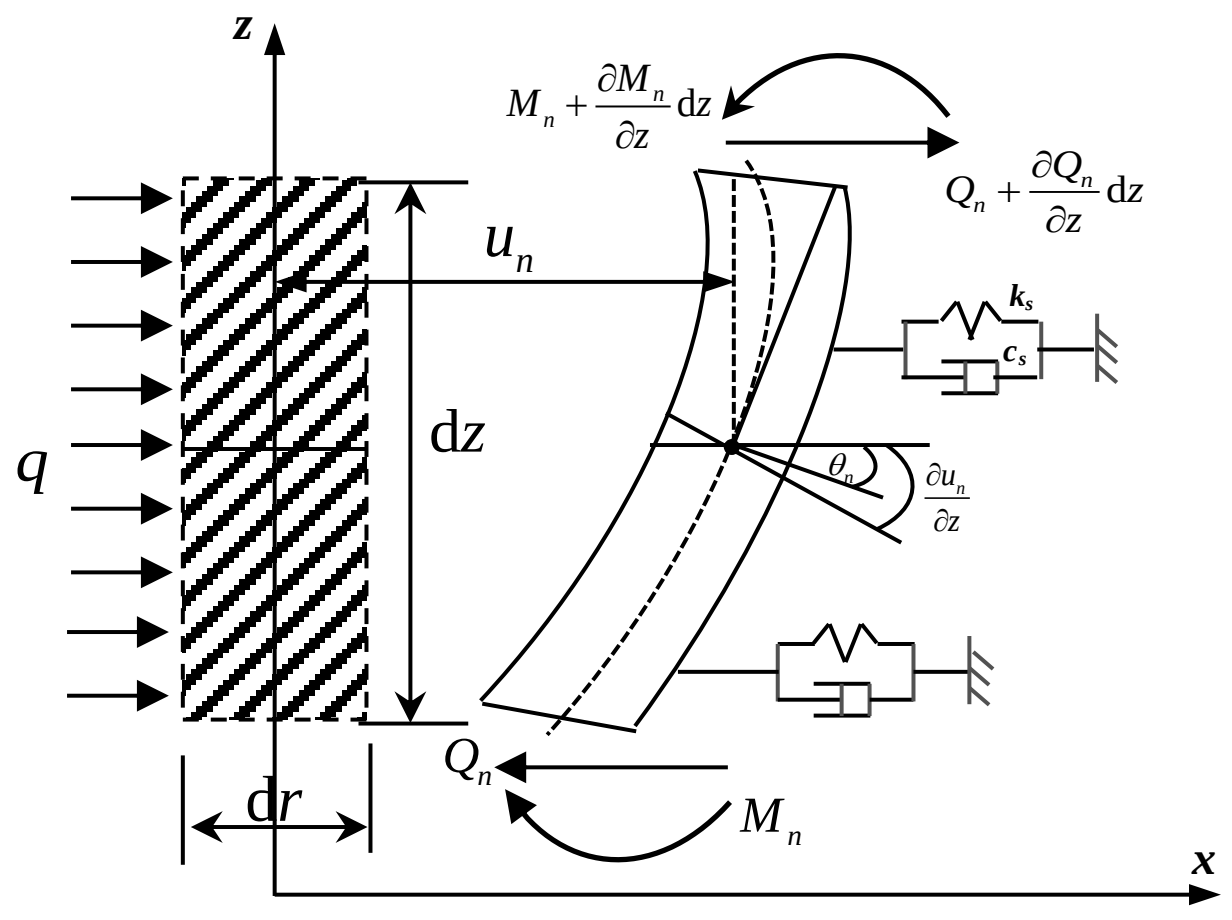

Fig.3 dynamic equilibrium of an infinitesimal pile element.

## 2.1 Governing equations and boundary conditions

(1) Governing equations of the row piles

The governing equations of the piles are expressed as follows [22-26]:

$$-\frac{\partial}{\partial z}\left[\kappa G_{\mathrm{p}} A_{\mathrm{p}}\left(\frac{\partial u_{mn}}{\partial z}-\theta_{mn}\right)\right]+\delta_{1n}\left(k_{\mathrm{s}} u_{mn}+c_{\mathrm{s}} \frac{\partial u_{mn}}{\partial t}\right)+\rho_{\mathrm{p}} A_{\mathrm{p}} \frac{\partial^{2} u_{mn}}{\partial t^{2}}=0 \tag{1}$$

$$\frac{\partial}{\partial z}\left(E_{\mathrm{p}} I_{\mathrm{p}} \frac{\partial \theta_{mn}}{\partial z}\right)+\kappa G_{\mathrm{p}} A_{\mathrm{p}}\left(\frac{\partial u_{mn}}{\partial z}-\theta_{mn}\right)-\rho_{\mathrm{p}} I_{\mathrm{p}}\left(\frac{\partial \theta_{mn}}{\partial t} \beta_{\mathrm{p}}^{*}+\frac{\partial^{2} \theta_{mn}}{\partial t^{2}}\right)=0 \tag{2}$$

where $E_{\mathrm{p}}$, $I_{\mathrm{p}}$, $G_{\mathrm{p}}$, $A_{\mathrm{p}}$, $\rho_{\mathrm{p}}$ are Young's modulus, the polar moment of inertia, the shear modulus, the cross-sectional area, and the density of the row piles' segment, respectively; $\kappa$ is a constant depending on the shape of the pile cross-section [29], and herein $\kappa=9/10$; $\beta_{\mathrm{p}}^{*}=\sqrt{\frac{\kappa G_{\mathrm{p}} A_{\mathrm{p}}}{\rho_{\mathrm{p}} I_{\mathrm{p}}}} \cdot \beta_{\mathrm{p}}$, $\beta_{\mathrm{p}}$ is the rotational damping coefficient [30]; $\delta_{1n}$ is Kronecker delta; $k_{\mathrm{s}}$ and $c_{\mathrm{s}}$ are the stiffness and the damping coefficients of the soil, whose empirical values are determined as [31]:

$$k_{\mathrm{s}}=1.2 E_{\mathrm{s}} \tag{3a}$$

$$c_{\mathrm{s}}=12\left(\frac{2 \omega r_{0}}{v_{\mathrm{s}}}\right)^{-1/4} r_{0} \rho_{\mathrm{s}} v_{\mathrm{s}}+2 \beta_{\mathrm{s}} \frac{k_{3}}{\omega} \tag{3b}$$

where $E_{\mathrm{s}}$, $\rho_{\mathrm{s}}$, $v_{\mathrm{s}}$, $\beta_{\mathrm{s}}$ are Young's modulus, the density, the shear wave velocity and the hysteretic damping coefficient of the soil, respectively; $r_0$ is the pile radius; $\omega$ is the angular frequency of the harmonic vibration.

(2) Governing equation of the sheet

The governing equation for the sheet is expressed as follows [32,33]:

$$E_b A_b \frac{\partial^2 w_b}{\partial x^2} + A_b \xi_b \frac{\partial^3 w_b}{\partial x^2 \partial t} - \rho_b A_b \frac{\partial^2 w_b}{\partial t^2} = 0 \tag{4}$$

where $E_b$, $A_b$, $\xi_b$ and $\rho_b$ are Young's modulus, the cross-sectional area, the damping coefficient and the density of the sheet, respectively; $w_b(z,t)$ is the longitudinal deflection.

## 2.2 Boundary conditions

For the lateral vibration of the piles, the neighboring pile segments satisfy the compatibility (displacements) and equilibrium (stresses) conditions at their interfaces:

$$u_{mn}(z,t) = u_{m(n+1)}(z,t)\Big|_{z=h_n} \tag{5.a}$$

$$\theta_{mn}(z,t) = \theta_{m(n+1)}(z,t)\Big|_{z=h_n} \tag{5.b}$$

$$\frac{\partial \theta_{mn}(z,t)}{\partial z} = \frac{\partial \theta_{m(n+1)}(z,t)}{\partial z}\Bigg|_{z=h_n} \tag{5.c}$$

$$\kappa G_p A_p \left[\frac{\partial u_{mn}(z,t)}{\partial z} - \theta_{mn}(z,t)\right] = \delta_{2n}\delta_{1m}Q(t) + \kappa G_p A_p \left[\frac{\partial u_{m(n+1)}(z,t)}{\partial z} - \theta_{m(n+1)}(z,t)\right]\Bigg|_{z=h_n} \tag{5.d}$$

where $h_n$ is the height of the pile segment $n$; $\delta_{2n}$ is Kronecker delta; $Q_b(t)$ denotes the lateral excitation; $m$=1 and 2 represent the front and rear row pile respectively, while $n$=1 and 2 denote the interfaces between the piles segments 1-3.

In addition, the boundary conditions of pile segments 1 and 3 for both the front and rear row piles are introduced as follows.

The toes of piles' segment 1are free:

$$\frac{\partial \theta_{m1}(z,t)}{\partial z}\Bigg|_{z=0} = 0 \tag{6.a}$$

$$\left.\frac{\partial u_{m1}(z,t)}{\partial z}-\theta_{m1}(z,t)\right|_{z=0}=0 \tag{6.b}$$

The top of the front pile's segment 3 are hinged:

$$\left.\frac{\partial \theta_{13}(z,t)}{\partial z}\right|_{z=h_3}=0 \tag{7.a}$$

$$u_{13}(z,t)\big|_{z=h_3}=w_{\mathrm{b}}(x,t)\big|_{x=0} \tag{7.b}$$

$$\kappa G_{\mathrm{p}}A_{\mathrm{p}}\left.\left(\frac{\partial u_{13}(z,t)}{\partial z}-\theta_{13}(z,t)\right)\right|_{z=h_3}=E_{\mathrm{b}}A_{\mathrm{b}}\left.\frac{\partial w_{\mathrm{b}}(x,t)}{\partial x}\right|_{x=0} \tag{7.c}$$

Similarly, the top of the rear row pile's segment 3 are hinged:

$$\left.\frac{\partial \theta_{23}(z,t)}{\partial z}\right|_{z=h_3}=0 \tag{8.a}$$

$$u_{23}(z,t)\big|_{z=h_3}=w_{\mathrm{b}}(x,t)\big|_{x=l} \tag{8.b}$$

$$\kappa G_{\mathrm{p}}A_{\mathrm{p}}\left.\left(\frac{\partial u_{23}(z,t)}{\partial z}-\theta_{23}(z,t)\right)\right|_{z=h_3}=E_{\mathrm{b}}A_{\mathrm{b}}\left.\frac{\partial w_{\mathrm{b}}(x,t)}{\partial x}\right|_{x=l} \tag{8.c}$$

## 3. Analytical solution

### 3.1 General solution

Considering lateral harmonic vibration of the pile segments, it is difficult to solve the differential equation from the governing equations. It is effective to transform the lateral displacements and rotations into a frequency domain equation, which can be formulated in the following form:

$$u(z,t)=U_{mn}(z,\omega)\mathrm{e}^{st} \tag{9 a}$$

$$\theta_{mn}(z,t)=\Theta_{mn}(z,\omega)\mathrm{e}^{st} \tag{9.b}$$

$$Q(t)=Q(\omega)\mathrm{e}^{st} \tag{9.c}$$

where $U_{mn}(z,\omega)$ and $\Theta_n(z,\omega)$ are the amplitude-frequency functions of the lateral displacement and the rotation due to the bending of the pile segment *n*, respectively; $Q(\omega)$ is the amplitude function of external excitation in frequency domain; $s = \mathrm{i}\omega$ and $\mathrm{i} = \sqrt{-1}$ is the imaginary unit.

The Laplace transform of Eq. (1) and (2) are expressed as:

$$-\kappa_{\mathrm{p}} G_{\mathrm{p}} A_{\mathrm{p}} \frac{\mathrm{d}^2 U_{mn}}{\mathrm{d}z^2} + \frac{\mathrm{d}\Theta_{mn}}{\mathrm{d}z} + \left[ s^2 \rho_{\mathrm{p}} A_{\mathrm{p}} + \delta_{1n} \left( k_{\mathrm{s}} - s c_{\mathrm{s}} \right) \right] U_{mn} = 0 \tag{10}$$

$$E_{\mathrm{p}} I_{\mathrm{p}} \frac{\mathrm{d}^2 \Theta_{mn}}{\mathrm{d}z^2} + \kappa_{\mathrm{p}} G_{\mathrm{p}} A_{\mathrm{p}} \frac{\mathrm{d}U_{mn}}{\mathrm{d}z} + \left[ \rho_{\mathrm{p}} I_{\mathrm{p}} \left( -s^2 + s\beta_{\mathrm{p}}^{*} \right) + s^2 \rho_{\mathrm{p}} A_{\mathrm{p}} + \delta_{1n} \left( k_{\mathrm{s}} - s c_{\mathrm{s}} \right) \right] \Theta_{mn} = 0 \tag{11}$$

In order to simplify the derivation procedure, let:

$$\begin{cases} W_{\mathrm{p}} = E_{\mathrm{p}} I_{\mathrm{p}} \\ J_{\mathrm{p}} = \kappa_{\mathrm{p}} G_{\mathrm{p}} A_{\mathrm{p}} \\ T_{\mathrm{p}} = \rho_{\mathrm{p}} I_{\mathrm{p}} \left( -s^2 + s\beta_{\mathrm{p}}^{*} \right) \\ K_{\mathrm{p}n} = s^2 \rho_{\mathrm{p}} A_{\mathrm{p}} + \delta_{1n} \left( k_{\mathrm{s}} - s c_{\mathrm{s}} \right) \end{cases} \tag{12}$$

From the above equations, the general solutions of the lateral displacement, the bending rotation angle, the shear force and the bending moment of the front pile segment ( $n = 1, 2, 3$ ) can be obtained as follows:

$$\begin{bmatrix} U_{mn} \\ \Theta_{mn} \\ Q_{mn}^{\mathrm{L}} \\ M_{mn}^{\mathrm{L}} \end{bmatrix} = \begin{bmatrix} T_{11}^{mn} & T_{12}^{mn} & T_{13}^{mn} & T_{14}^{mn} \\ T_{21}^{mn} & T_{22}^{mn} & T_{23}^{mn} & T_{24}^{mn} \\ T_{31}^{mn} & T_{32}^{mn} & T_{33}^{mn} & T_{34}^{mn} \\ T_{41}^{mn} & T_{42}^{mn} & T_{43}^{mn} & T_{44}^{mn} \end{bmatrix} \begin{bmatrix} A_{mn} \\ B_{mn} \\ C_{mn} \\ D_{mn} \end{bmatrix} \tag{13}$$

where $Q_{mn}^{\mathrm{L}}, M_{mn}^{\mathrm{L}}$ denotes the frequency domain formulation of the shear force and the moment of the pile, respectively. $A_{mn}$ , $B_{mn}$ , $C_{mn}$ , $D_{mn}$ are constant factors, which can be determined as shown in the following section. Coefficient matrix can be attained as detailed in the AppendixⅠ.

The Laplace transform of Eq. (4) is expressed as:

$$\left(E_{\mathrm{b}}A_{\mathrm{b}}+\mathrm{i}\omega A_{\mathrm{b}}\xi_{\mathrm{b}}W_{\mathrm{b}}\right)\frac{\mathrm{d}^2W_{\mathrm{b}}}{\mathrm{d}x^2}+\omega^2\rho_{\mathrm{b}}A_{\mathrm{b}}W_{\mathrm{b}}=0 \tag{14}$$

The general solutions of the horizontal displacement and axial force can be obtained as follows:

$$\begin{bmatrix} W_{\mathrm{b}} \\ N_{\mathrm{b}}^{\mathrm{L}} \end{bmatrix}=\begin{bmatrix} G_{11} & G_{12} \\ G_{21} & G_{22} \end{bmatrix}\begin{bmatrix} \ell_b \\ m_{\mathrm{b}} \end{bmatrix} \tag{15}$$

where $W_{\mathrm{b}}=W_{\mathrm{b}}\left(z,\omega\right)$, $N_{\mathrm{b}}^{\mathrm{L}}=N_{\mathrm{b}}^{\mathrm{L}}\left(z,\omega\right)$ denotes the frequency domain transform of the longitudinal deflection and the axial force of the sheet. The coefficient matrix is derived as detailed in the Appendix Ⅱ.

### 3.2 Matrix equation for the motion of sheet-pile groin

In order to solve the undetermined functions, the boundary conditions in the frequency domain should be taken into consideration as follows.

$$U_{mn}(z,\omega)=U_{m(n+1)}(z,\omega)\Big|_{z=h_n} \tag{16.a}$$

$$\Theta_{mn}(z,\omega)=\Theta_{m(n+1)}(z,\omega)\Big|_{z=h_n} \tag{16.b}$$

$$\frac{\mathrm{d}\Theta_{mn}(z,\omega)}{\mathrm{d}z}=\frac{\mathrm{d}\Theta_{m(n+1)}(z,\omega)}{\mathrm{d}z}\Bigg|_{z=h} \tag{16.c}$$

$$\kappa G_{\mathrm{p}}A_{\mathrm{p}}\left[\frac{\mathrm{d}U_{mn}(z,\omega)}{\mathrm{d}z}-\Theta_{mn}(z,\omega)\right]=\delta_{2n}Q(t)+\kappa G_{\mathrm{p}}A_{\mathrm{p}}\left[\frac{\mathrm{d}U_{m(n+1)}(z,\omega)}{\mathrm{d}z}-\Theta_{m(n+1)}(z,\omega)\right]\Bigg|_{z=h_n} \tag{16.d}$$

And the following matrix can be established:

$$\begin{bmatrix} T_{11}^{mn} & T_{21}^{mn} & T_{31}^{mn} & T_{41}^{mn} \\ T_{12}^{mn} & T_{22}^{mn} & T_{32}^{mn} & T_{42}^{mn} \\ T_{13}^{mn} & T_{23}^{mn} & T_{33}^{mn} & T_{43}^{mn} \\ T_{14}^{mn} & T_{24}^{mn} & T_{34}^{mn} & T_{44}^{mn} \\ -T_{11}^{m(n+1)} & -T_{21}^{m(n+1)} & -T_{31}^{m(n+1)} & -T_{41}^{m(n+1)} \\ -T_{12}^{m(n+1)} & -T_{22}^{m(n+1)} & -T_{32}^{m(n+1)} & -T_{42}^{m(n+1)} \\ -T_{13}^{m(n+1)} & -T_{23}^{m(n+1)} & -T_{33}^{m(n+1)} & -T_{43}^{m(n+1)} \\ -T_{14}^{m(n+1)} & -T_{24}^{m(n+1)} & -T_{34}^{m(n+1)} & -T_{44}^{m(n+1)} \end{bmatrix}^{\mathrm{T}} \begin{bmatrix} A_{mn} \\ B_{mn} \\ C_{mn} \\ D_{mn} \\ A_{m(n+1)} \\ B_{m(n+1)} \\ C_{m(n+1)} \\ D_{m(n+1)} \end{bmatrix} = \begin{bmatrix} 0 \\ 0 \\ \delta_{2n}\delta_{1m}Q_{\mathrm{b}} \\ 0 \end{bmatrix} \tag{17}$$

According to Eqs. (6.a-6.b), its frequency domain expression can be written as:

$$\left.\frac{\mathrm{d}\Theta_{mn}(z,\omega)}{\mathrm{d}z}\right|_{z=0} = 0 \tag{18.a}$$

$$\left.\frac{\mathrm{d}U_{mn}(z,\omega)}{\mathrm{d}z} - \Theta_{mn}(z,\omega)\right|_{z=0} = 0 \tag{18.b}$$

The following matrix can be established at $z = 0$:

$$\begin{bmatrix} T_{31}^{m1} & T_{41}^{m1} \\ T_{32}^{m1} & T_{42}^{m1} \\ T_{33}^{m1} & T_{43}^{m1} \\ T_{34}^{m1} & T_{44}^{m1} \end{bmatrix}^{\mathrm{T}} \begin{bmatrix} A_{m1} \\ B_{m1} \\ C_{mn1} \\ D_{m1} \end{bmatrix} = \begin{bmatrix} 0 \\ 0 \end{bmatrix} \tag{19}$$

According to Eq. (7.a) and Eq. (8.a), the following matrix can be established at $h_3$:

$$\begin{bmatrix} T_{41}^{m3} \\ T_{42}^{m3} \\ T_{43}^{m3} \\ T_{44}^{m3} \end{bmatrix}^{\mathrm{T}} \begin{bmatrix} A_{m3} \\ B_{m3} \\ C_{m3} \\ D_{m3} \end{bmatrix} = [0] \tag{20}$$

According to Eqs. (7.b-7.c) and Eqs. (8.b-8.c), the following matrix can be established:

$$\left.\begin{bmatrix} T_{11}^{13} & T_{31}^{13} \\ T_{12}^{13} & T_{32}^{13} \\ T_{13}^{13} & T_{33}^{13} \\ T_{14}^{13} & T_{34}^{13} \end{bmatrix}^{\mathrm{T}} \begin{bmatrix} A_{13} \\ B_{13} \\ C_{13} \\ D_{13} \end{bmatrix}\right|_{z=h_{\mathrm{p}}} - \left.\begin{bmatrix} G_{11} & G_{21} \\ G_{12} & G_{22} \end{bmatrix}^{\mathrm{T}} \begin{bmatrix} \ell_b \\ m_{\mathrm{b}} \end{bmatrix}\right|_{x=0} = \begin{bmatrix} 0 \\ 0 \end{bmatrix} \tag{21}$$

$$\begin{bmatrix} T_{11}^{23} & T_{31}^{23} \\ T_{12}^{23} & T_{32}^{23} \\ T_{13}^{23} & T_{33}^{23} \\ T_{14}^{23} & T_{34}^{23} \end{bmatrix}^{\mathrm{T}} \begin{bmatrix} A_{23} \\ B_{23} \\ C_{23} \\ D_{23} \end{bmatrix}\Bigg|_{z=h_{\mathrm{p}}} - \begin{bmatrix} G_{11} & G_{21} \\ G_{12} & G_{22} \end{bmatrix}^{\mathrm{T}} \begin{bmatrix} \ell_b \\ m_{\mathrm{b}} \end{bmatrix}\Bigg|_{x=0} = \begin{bmatrix} 0 \\ 0 \end{bmatrix} \tag{22}$$

Accordingly, the undetermined constant factors of each pile segments and sheet can be solved by Cramer's rule or with the aid of matrix calculator.

### 3.3 Semi-analytical solution

The lateral excitation imposed on front row pile is idealized as a semi-sine excitation[34]:

$$Q(t) = Q_{\max} \sin\left(\theta t \cdot H\left(\frac{\pi}{\theta} - t\right)\right) \tag{23}$$

where $Q_{\max}$ and $\theta$ indicate the amplitude and angular frequency of the excitation, respectively; $H(\cdot)$ denotes the Heaviside step function. The excitation curve is depicted in Fig.4.

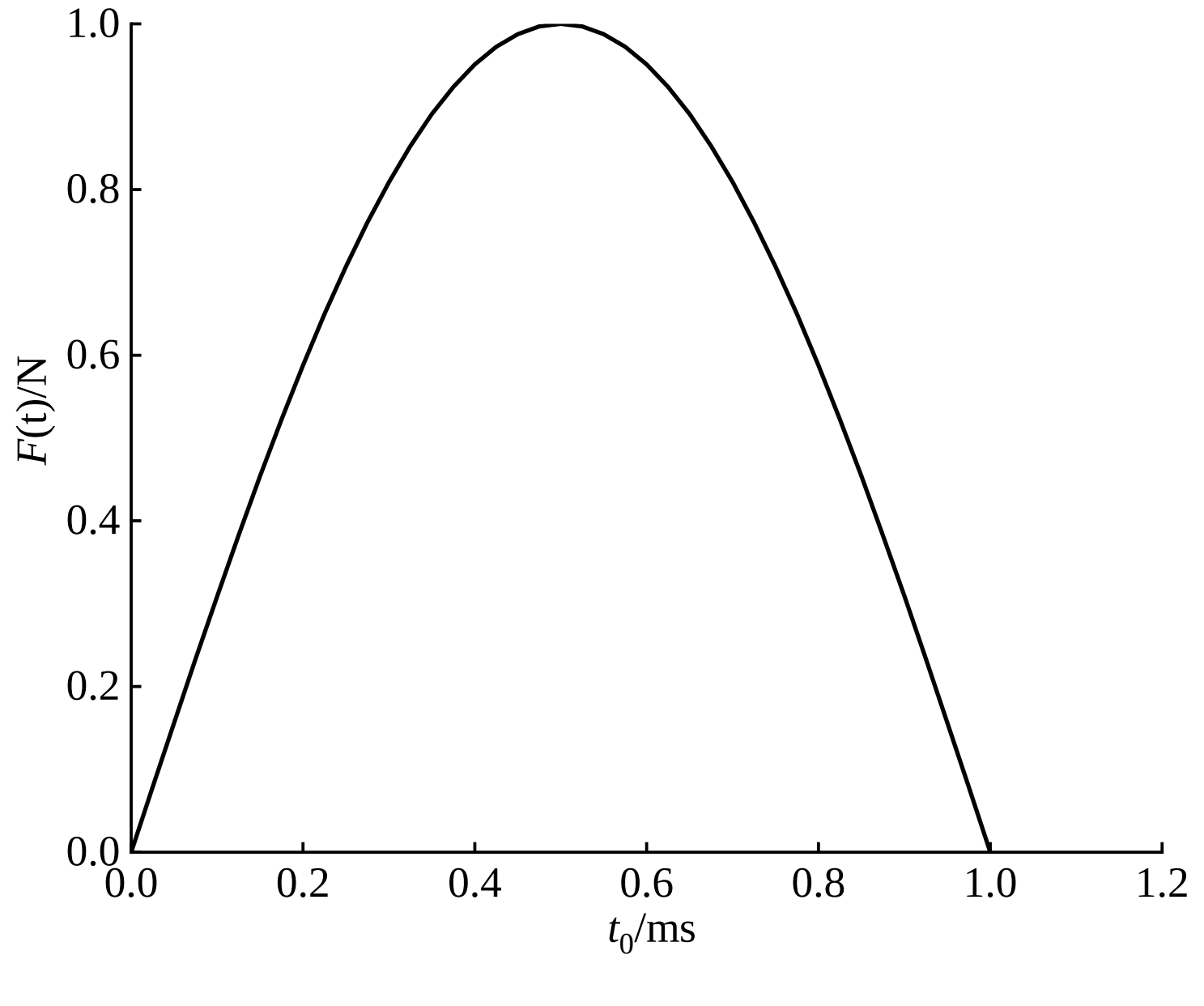

Fig.4 The semi-sine excitation curve

Accordingly, the dynamic reactions of a sheet-pile at arbitrary location subjected to semi-sine excitation in frequency domain have been obtained. Then, the analytical solution in time domain can be obtained by inverse Fourier transform as follows[33–35]:

$$u_{mp}(z,t) = \mathrm{IFT}\left[\frac{Q_{\max}\theta}{\theta^2 - \omega^2}\left(1 + \mathrm{e}^{-\mathrm{i}\omega\frac{\pi}{\theta}}\right)U_{mn}\left(z,\omega\right)\right] \tag{24}$$

$$v_{mp}(z,t) = \mathrm{IFT}\left[\frac{Q_{\max}\theta}{\theta^2 - \omega^2}\left(1 + \mathrm{e}^{-\mathrm{i}\omega\frac{\pi}{\theta}}\right)U_{mn}\left(z,\omega\right)\cdot \mathrm{i}\omega\right] \tag{25}$$

where $u_{mp}\left(z,t\right)$, $v_{mp}\left(z,t\right)$ is the lateral displacement and velocity, respectively.

## 4. Validation

Finite element analysis using ABAQUS is employed to validate the accuracy of the presented model in an intuitive way. The numerical model of a specific case is depicted in Fig.5. The sheet-pile groins are simulated by shear flexible beam elements (B21) that corresponds to TB and the surrounding soil is simulated by four-node plane strain elements (CPE4R). As shown in Fig.5, the extent of the soil model is 30 times the pile radius in the radial direction, and 3 times the embedded pile length in the vertical direction. The rotation of the pile top is fixed (Eq. (7.a)), and the pile-soil interaction is simulated by the embedded region constraint. The parameters of the pile and the soil are listed in Table 1. The model subjected to the lateral excitation is analyzed using the explicit solver of ABAQUS.

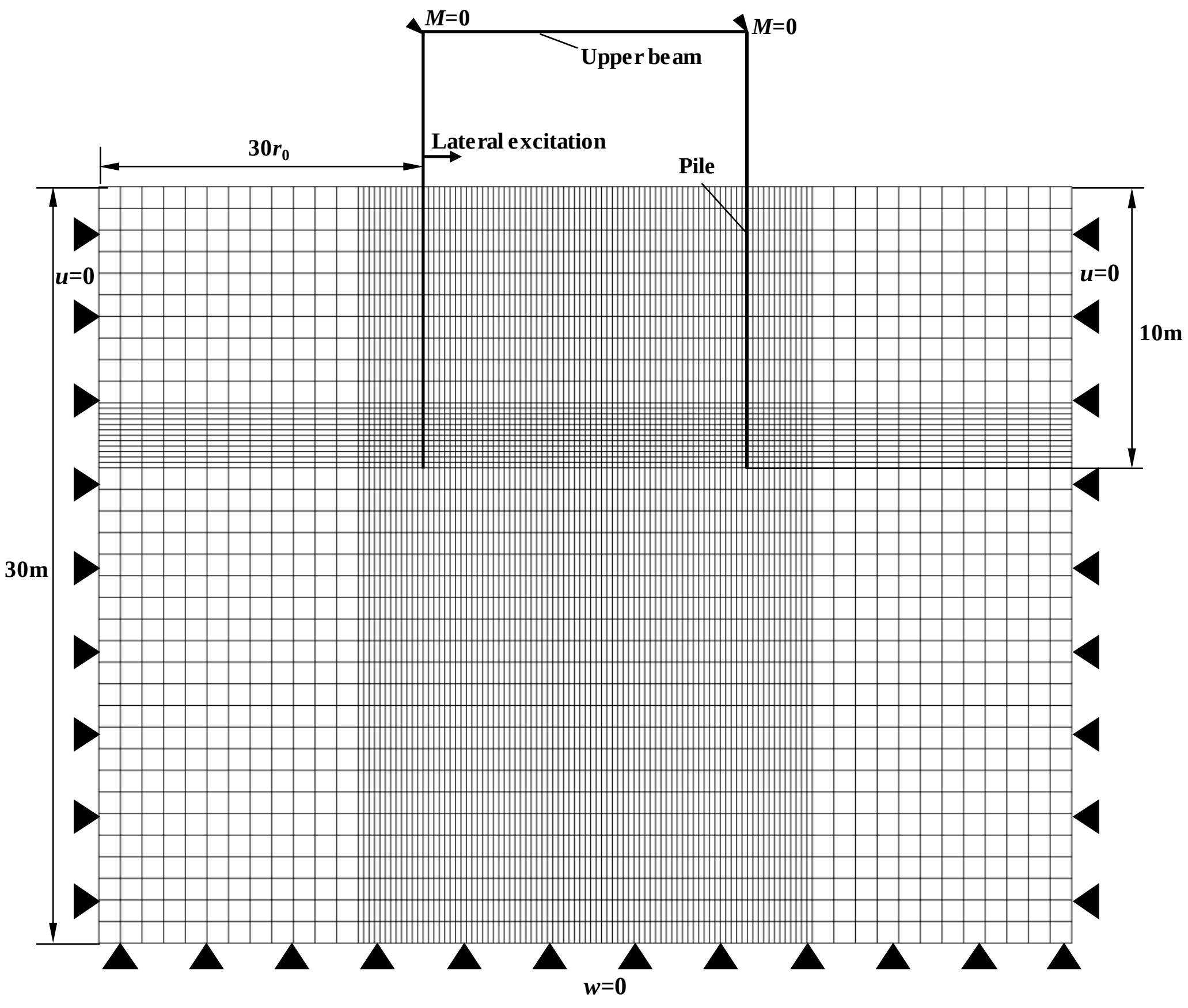

Fig.5. Schematic of the numerical model

Table 1 Properties of sheet-pile and soil materials.

| Material | Density | Young's modulus | Poisson's ratio | Damping | Shear wave velocity |
|---|---|---|---|---|---|
| Sheet-Pile | $\rho_p = 2500\text{kg/m}^3$ | $E_p = 40\text{GPa}$ | $\nu_p = 0.12$ | $\beta_p = 0.0001$ | \ |
| Soil | $\rho_s = 1800\text{kg/m}^3$ | $E_s = 40\text{MPa}$ | $\nu_s = 0.4$ | $\beta_s = 0.02$ | $v_s = 100\text{m/s}$ |
| $r_0 = 0.5\text{m}$, $h_1 = 10.0\text{m}$, $h_2 = 0.5\text{m}$, $h_3 = 5.0\text{m}$, $\pi/\theta T_c = 0.5$ | | | | | |

Fig.6 and Fig.7 compare the dynamic responses of a sheet-pile groin obtained from the developed solution and FEM simulation. In order to validate the velocity responses at different positions which is numbered and represented by the bold black dots in the vignette on the left, the dimensionless time and velocity are introduced as follows, i.e.:

$$\bar{t} = \frac{t}{T_c}, \quad T_c = \frac{h_1}{C_s} \tag{26}$$

$$\bar{v}_{mn} = v_{mn} \Big/ \frac{C_s}{J_p}, \quad C_s = \sqrt{\kappa G_p / \rho_p} \tag{27}$$

where $C_s$ is the one-dimensional shear wave velocity of the piles. $T_c$ is the defined propagation time for the intermediate variable of dimensionless time.

In order to evaluate the dynamic response of the sheet-pile groin when the excitation is acted, the numbered receiving points are set at different positions on the sheet and the row piles to study the vibration mechanism in Fig.6. It can be noted that the results of the analytical solution agree well with the FEA results in this case, especially the arrival time of the crest positions and the amplitude of the velocity. On the whole, more obvious reflection of the crest can be seen in the first two cycles in Fig.7, and then gradually stabilized due to wave dissipation and applied damping. The interpretation of the velocity response curve will be detailed in the ensuing parametric study.

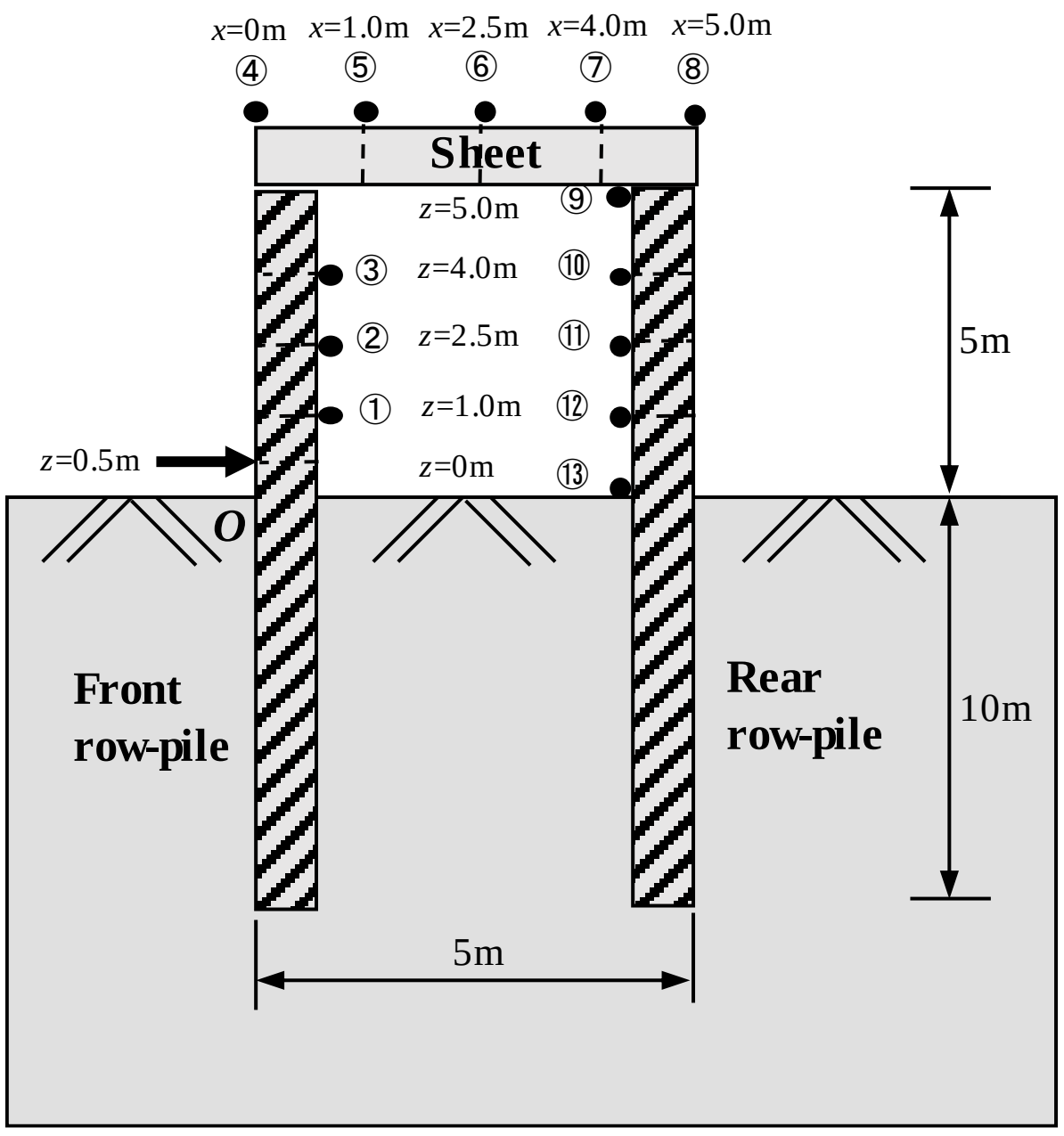

Fig.6. The sheet-pile groin with numbered receiving points under lateral excitation

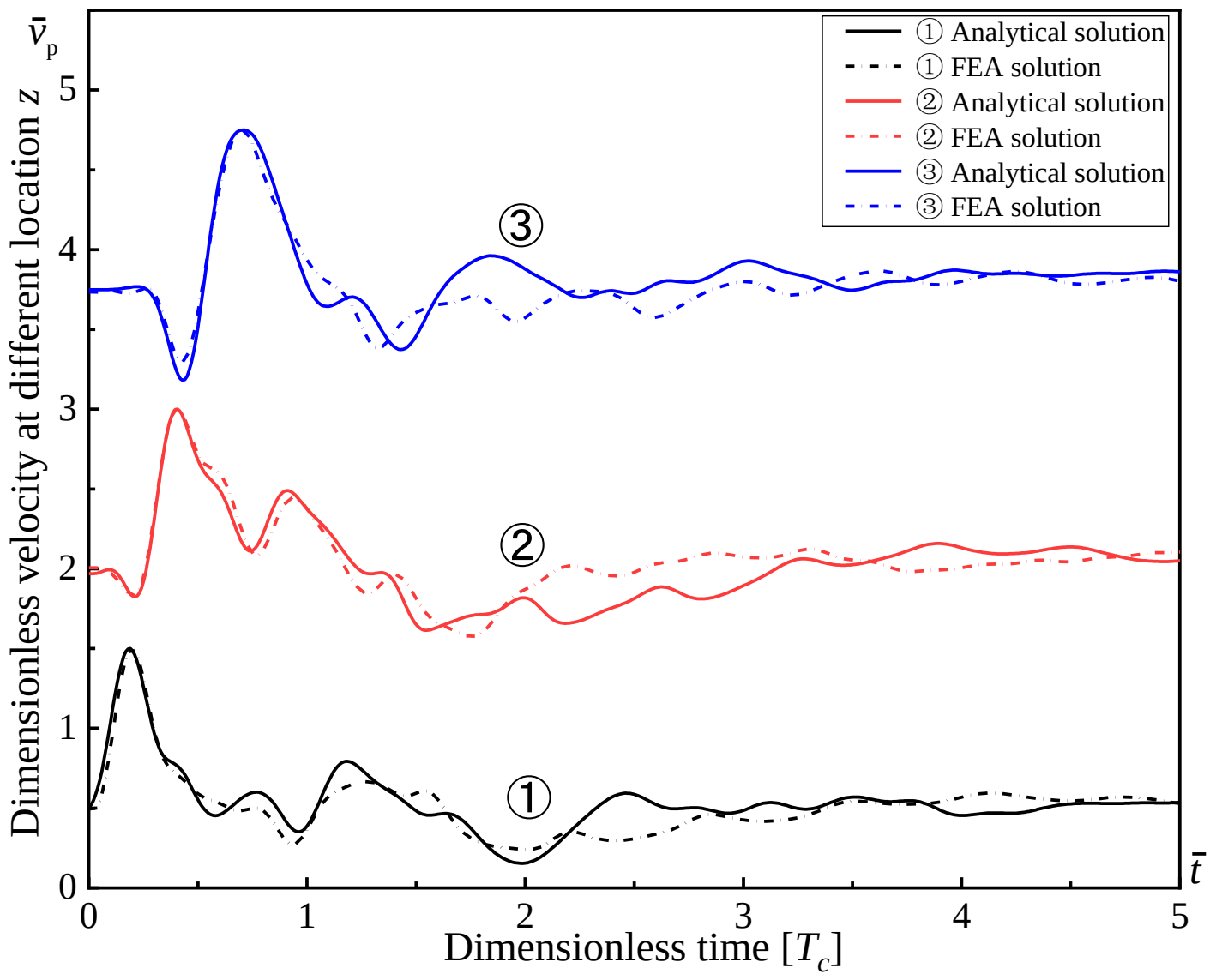

Fig.7a. The front row pile with lateral excitation imposed at: $z = 1.0m, 2.5m, 4.0m$

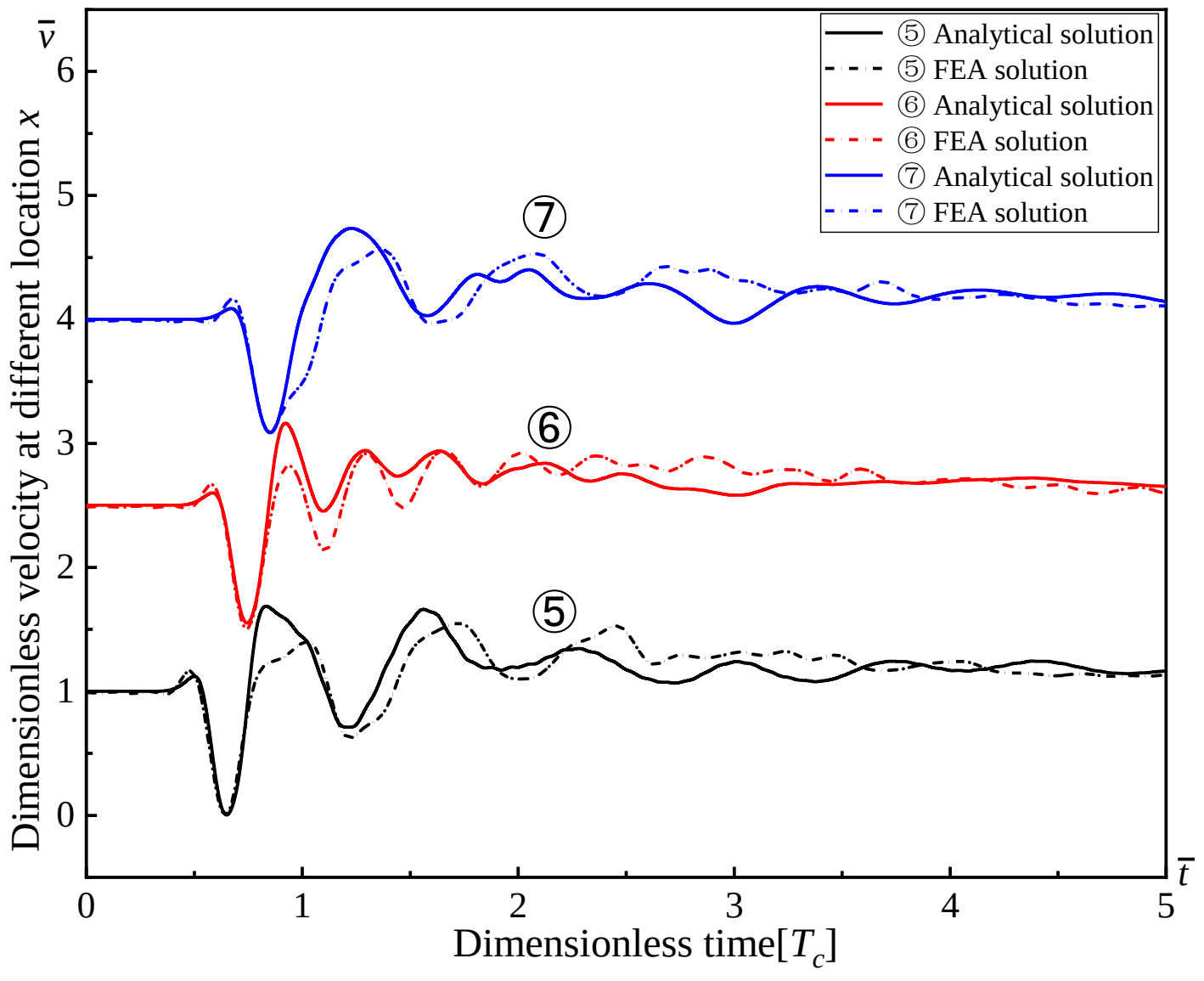

Fig.7. The sheet imposed at： $x = 1.0m, 2.5m, 4.0m$ in time domain

## 5. Parametric study and discussion

It can be seen from Fig.8 and Fig.9 that the regularities of the sheet and the rear row pile are generally similar, and the arrival time is different. In addition, it can be clearly found that the velocity amplitude of the end of the sheet and the top of the rear row pile, i.e., the joint of the sheet and rear row pile, is significantly greater than other positions. Because more attention is paid to the maximum velocity amplitude of dynamic response in practice, subsequent research only focuses on the vibration at the joint of the sheet and the rear row pile.

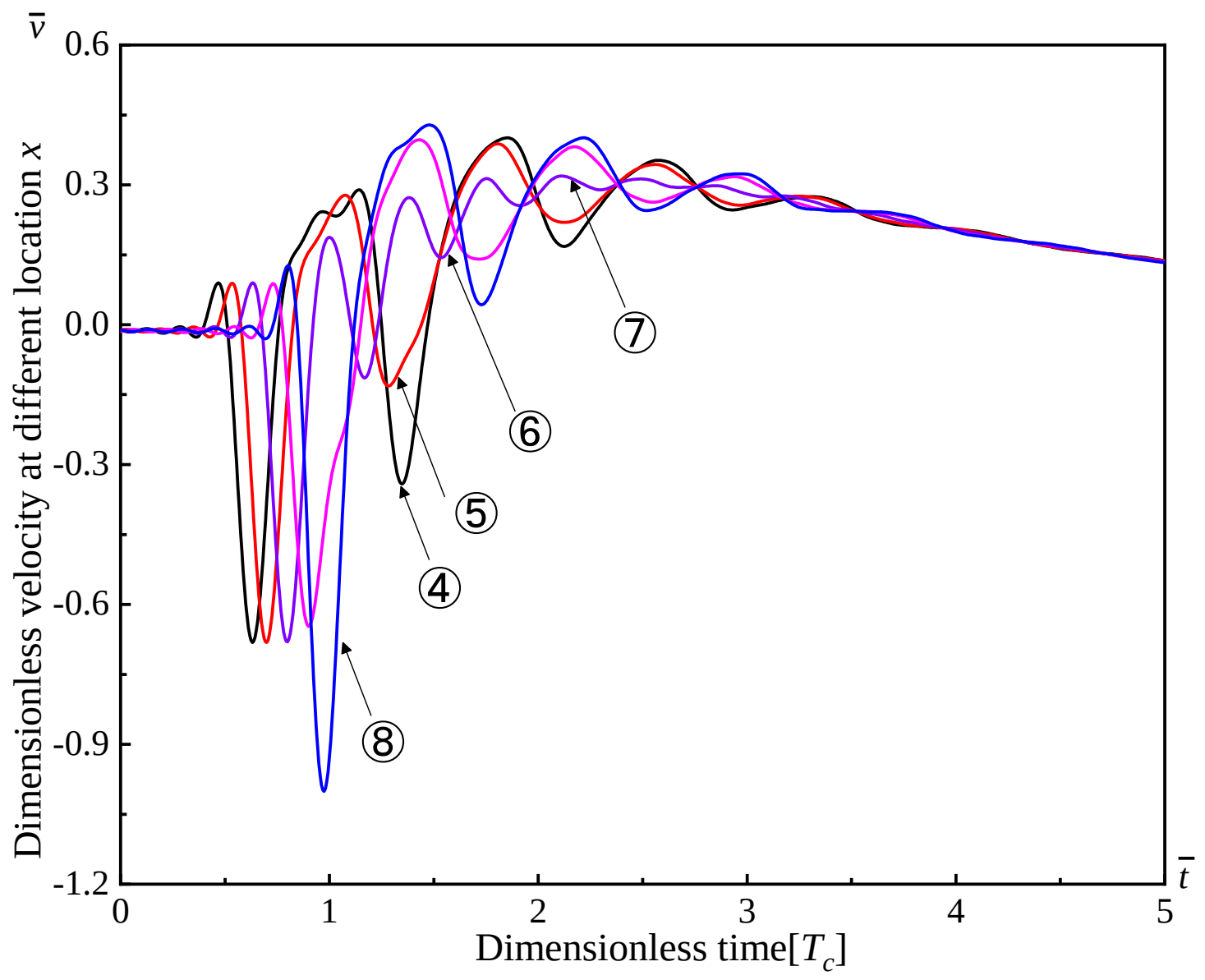

Fig.8 Dynamic response of the sheet at different location $x$

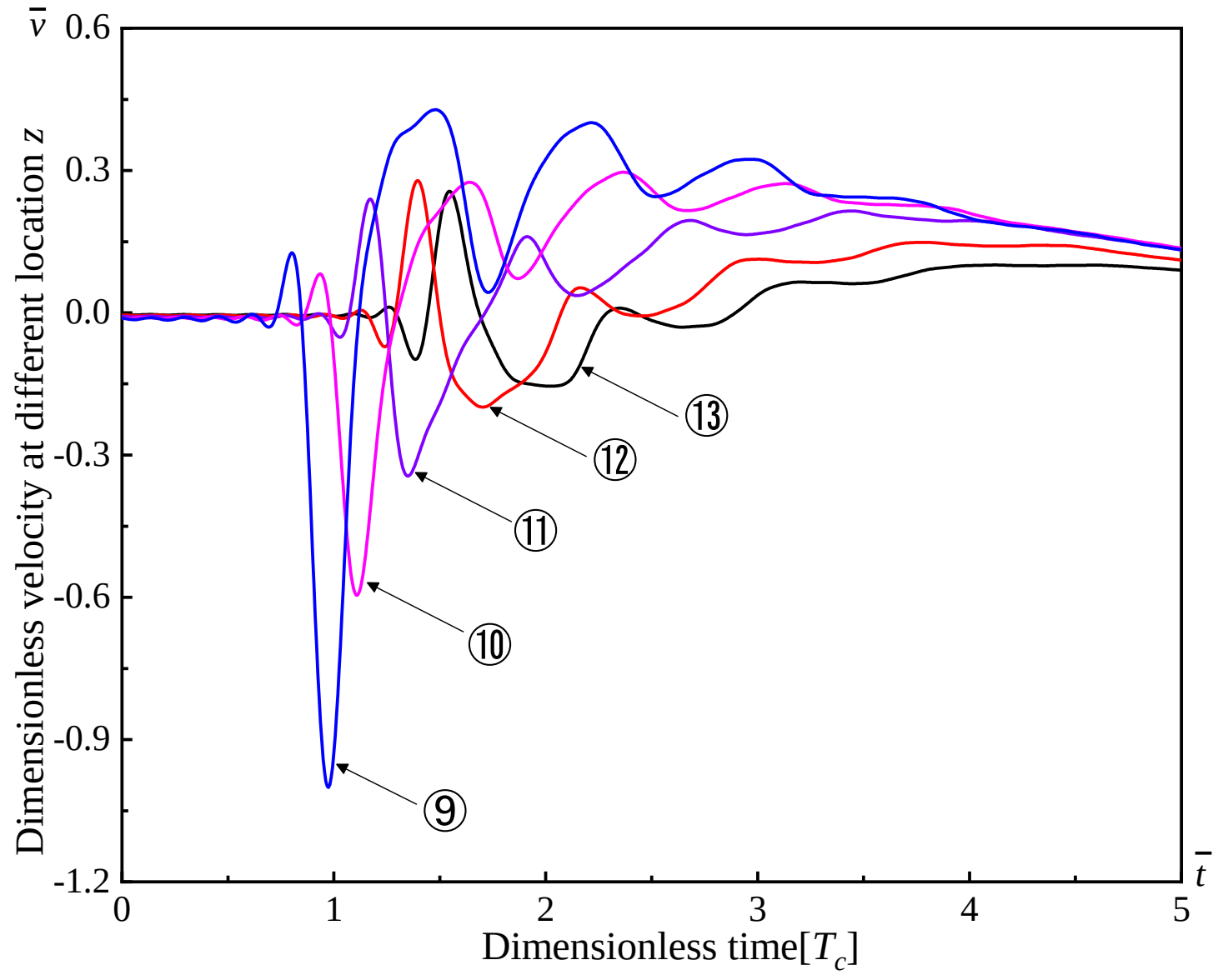

Fig.9 Dynamic response of the rear row pile at different location $z$

Having determined the maximum response position, the factors that influence the dynamic response of the sheet-pile groin are analyzed below. If not mentioned otherwise, the default parameters in this section are given in Table 1.

### 5.1 Radius of the piles

Fig.10 and Fig.11 demonstrate the effect of the piles radius on dynamic response of the sheet-pile groin in time and frequency domain, respectively. The velocity amplitude becomes greater for smaller pile radius in Fig.10. It is also worth noting that a larger radius corresponds to a faster propagating wave velocity that has been interpreted by Wu et al.[25], which accounts for the phenomenon that the first incident wave is in advanced in a pile with larger radius. The amplitude-frequency curve gradually becomes disordered as the frequency increases in Fig.11. From the picture that was studied most in low-frequency resonance, it can be found that a larger pile radius corresponds to a greater resonance frequency. It can be interpreted that the rigidity of the pile will increase as the pile radius increases, which makes the resonance of the structure more difficult to occur. And it also can be found that the greater velocity amplitude corresponds to the smaller pile radius. This can be understood that the overall stiffness of the sheet-pile structure is enhanced, which will limit the movement of the joint of the sheet-pile structure.

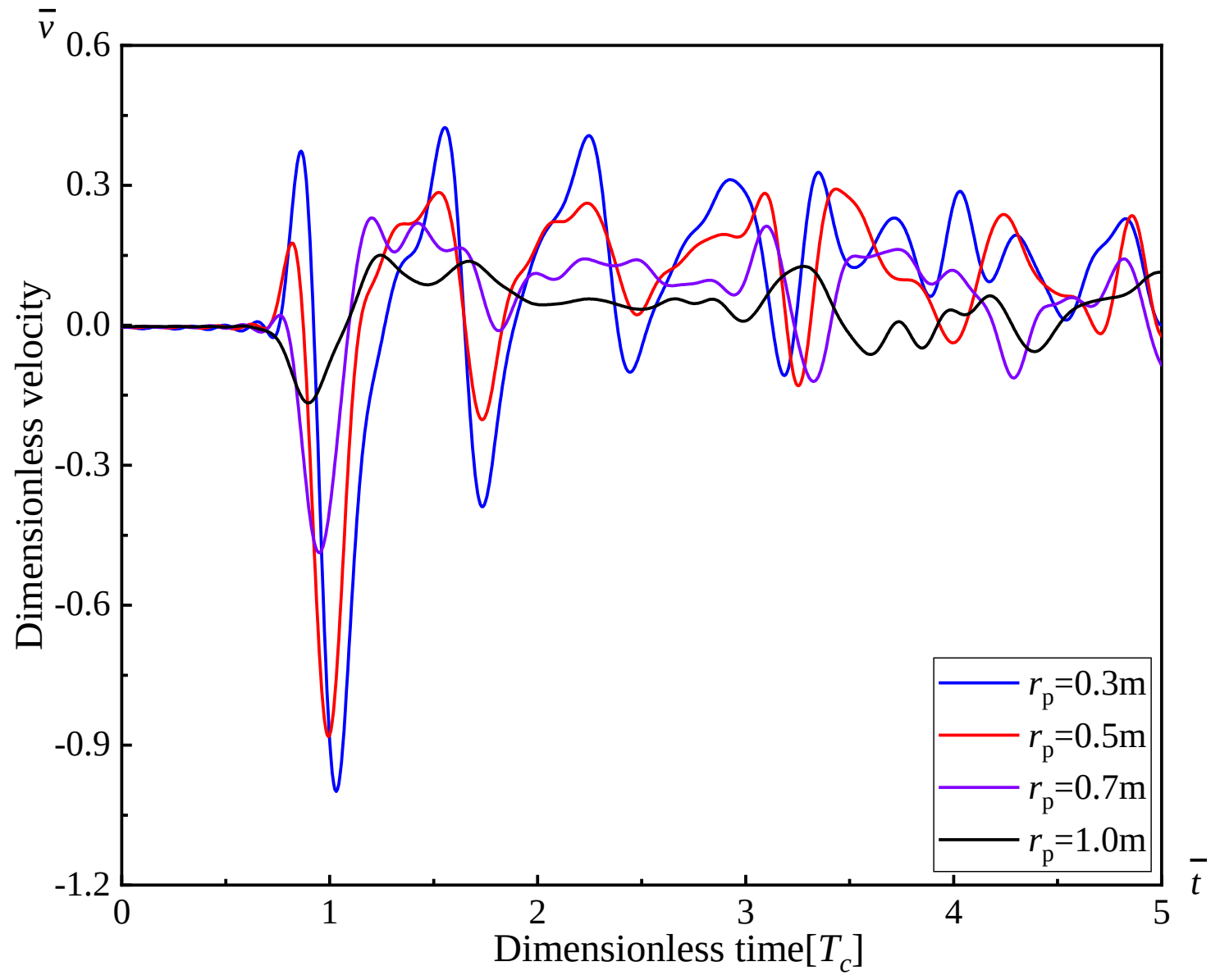

Fig.10. Effect of the piles radius on the velocity in time domain

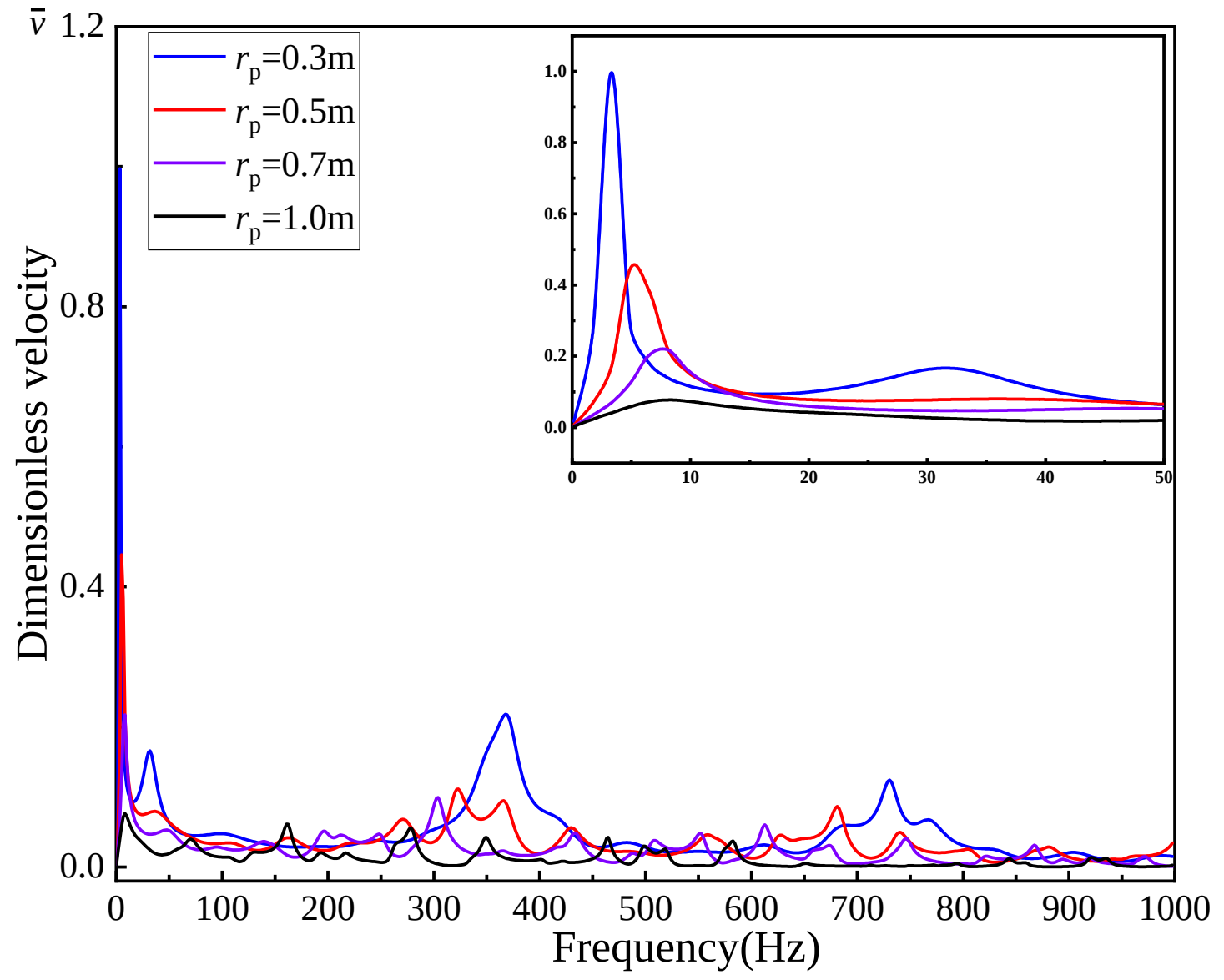

Fig.11. Effect of the pile radius on the velocity in frequency domain

## 5.2 Embedded ratio of the piles

Fig.12 and Fig.13 demonstrate the effect of the embedded ratio of the piles on dynamic response in time and frequency domain, respectively. The incident wave reaches the peak first and the velocity amplitude becomes greater for smaller embedded ratio of the piles in Fig.12. This is because a shorter embedding ratio corresponds to a shorter pile length above the ground under the same conditions. It needs less propagation time for the joint to receive excitation and produces stronger effect on the receiving velocity waves. When the embedded ratio increases, the dynamic response of the sheet-pile structure is approximately manifested as the vibration of the single pile itself. The amplitude-frequency curve is depicted in Fig.13. The greater embedded ratio corresponds to a smaller resonance frequency. It can be illustrated that the vibration mode of the pile is more likely to change and the natural frequency decreases when the embedding ratio of the pile increases, i.e., resonance is more likely to occur. The velocity amplitude first increases and then decreases as the embedding ratio increases in low-frequency resonance. It can be explained that the rigidity of the structure decreases and the vibration effect of the velocity amplitude on the joint gradually increases as the pile becomes longer. However, when the pile continues to becomes longer, the response of the sheet-pile groin is gradually appeared as the single pile and the restraint effect of the sheet gradually weakens.

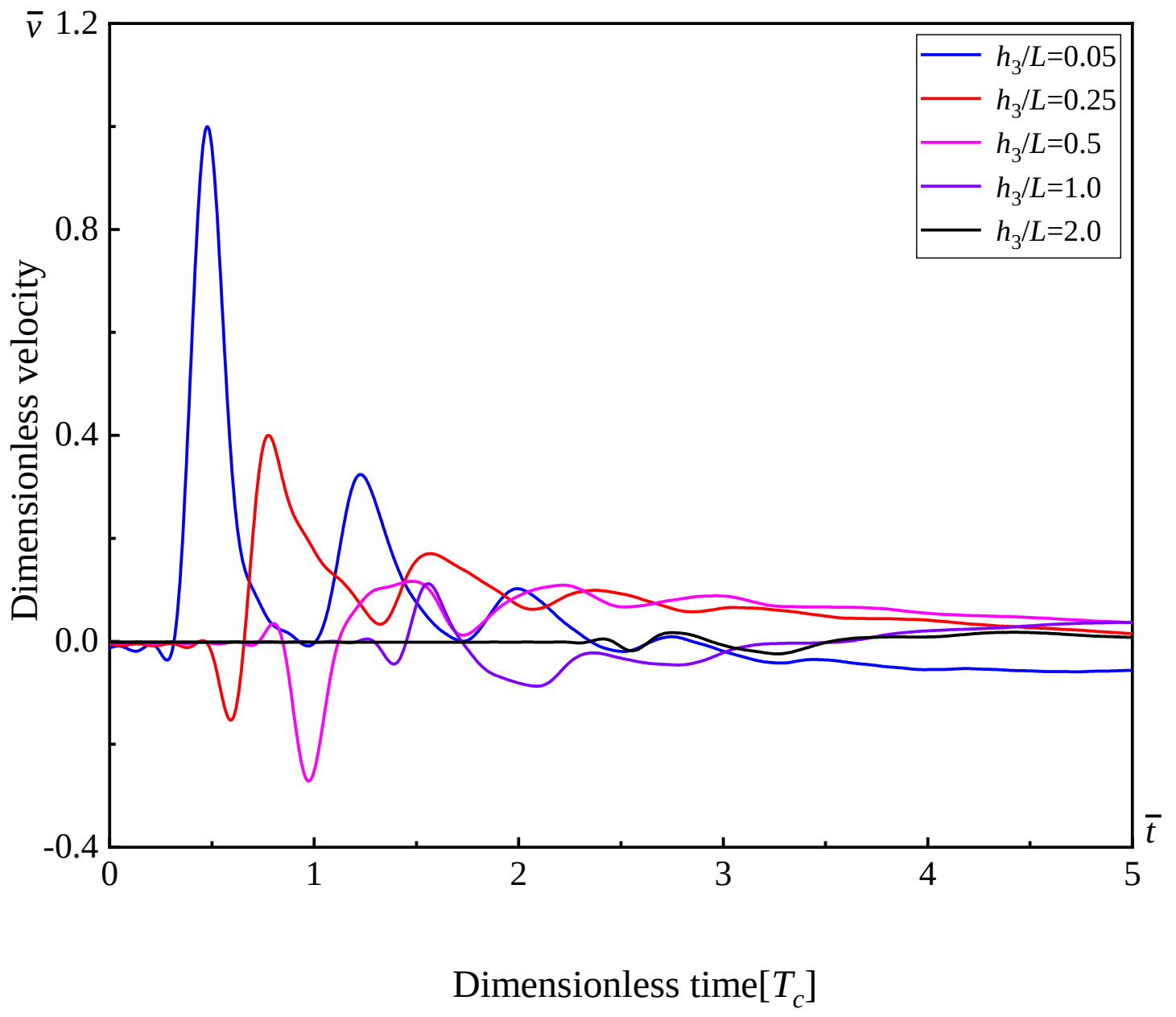

Fig.12. Effect of the embedded ratio of the piles on the velocity in time domain

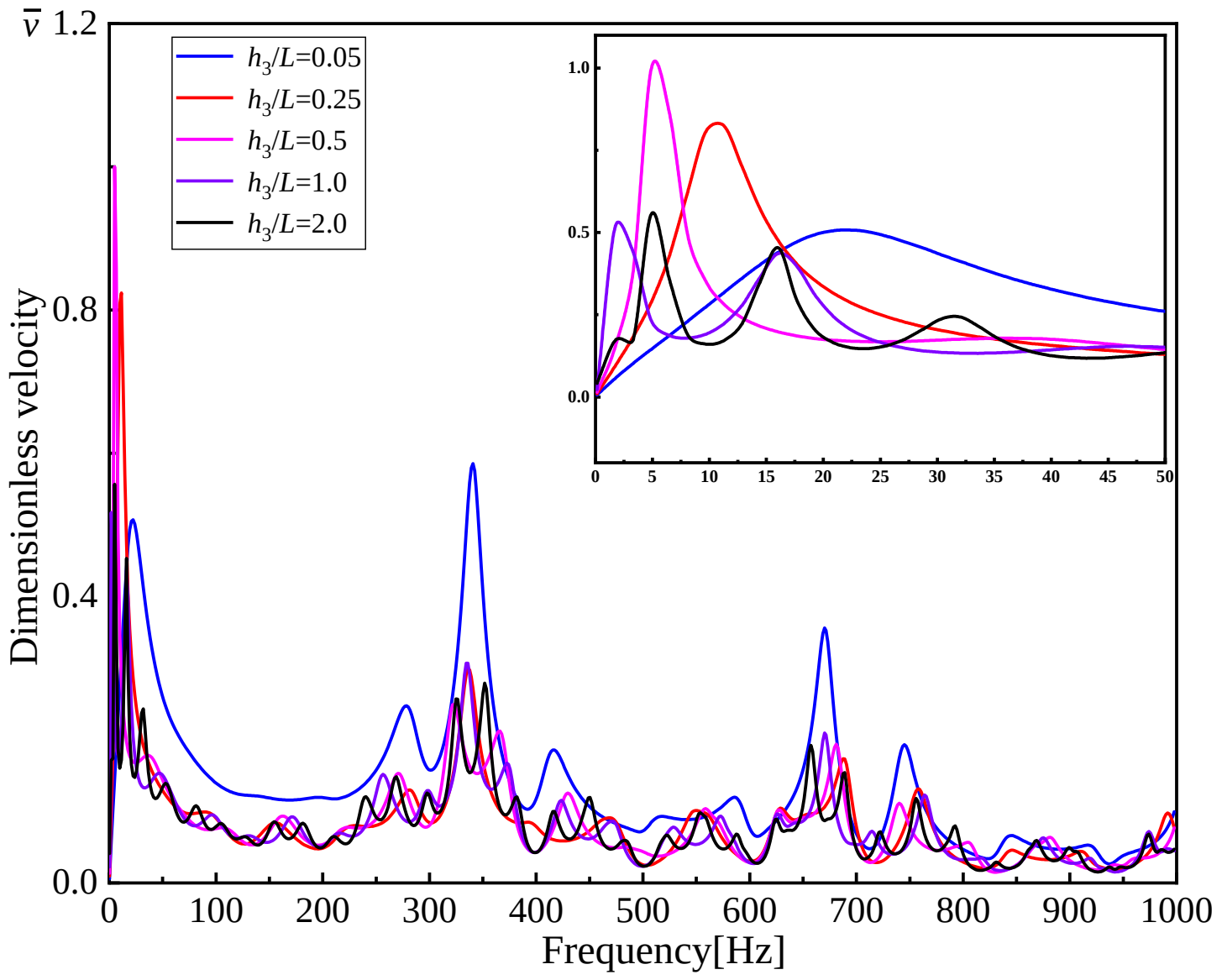

Fig.13. Effect of the embedded ratio of the piles on the velocity in frequency domain

## 5.3 Radius of the sheet

Fig.14 and Fig.15 show the effect of the sheet radius on dynamic response in time and frequency domain, respectively. From the Fig.14, it can be found that the incident wave reaches the first crest almost simultaneously. This is because the sheet is assumed to be a one-dimensional rod. When the sheet radius increases, the propagation speed of the elastic wave is equal to the compression wave velocity without changing, so the arrival time of the first crest is the same. And the maximum velocity amplitude corresponds to a smaller sheet radius in Fig.14.It can be interpreted that the increase of the compressive rigidity of the sheet enhances the restraining effect on the movements of the sheet, so a larger velocity amplitude corresponds to a smaller sheet radius. The amplitude-frequency curve gradually approaches at high frequencies as the sheet radius increases in Fig.15. The greater sheet radius corresponds to a smaller resonance frequency. It can be explained that the sheet radius influences the restraint methods of the joints which is usually assumed to be a spring constraint. When the sheet radius increases, the end restraint increases, resulting in the reduction of the natural frequency of the sheet. The specific validation and interpretation process can be found in [36].The velocity amplitude decreases as the sheet radius increases in low-frequency resonance. It can be interpreted that the compressive rigidity of the sheet enhances the restraining effect on the movements of the sheet as the sheet radius increases...

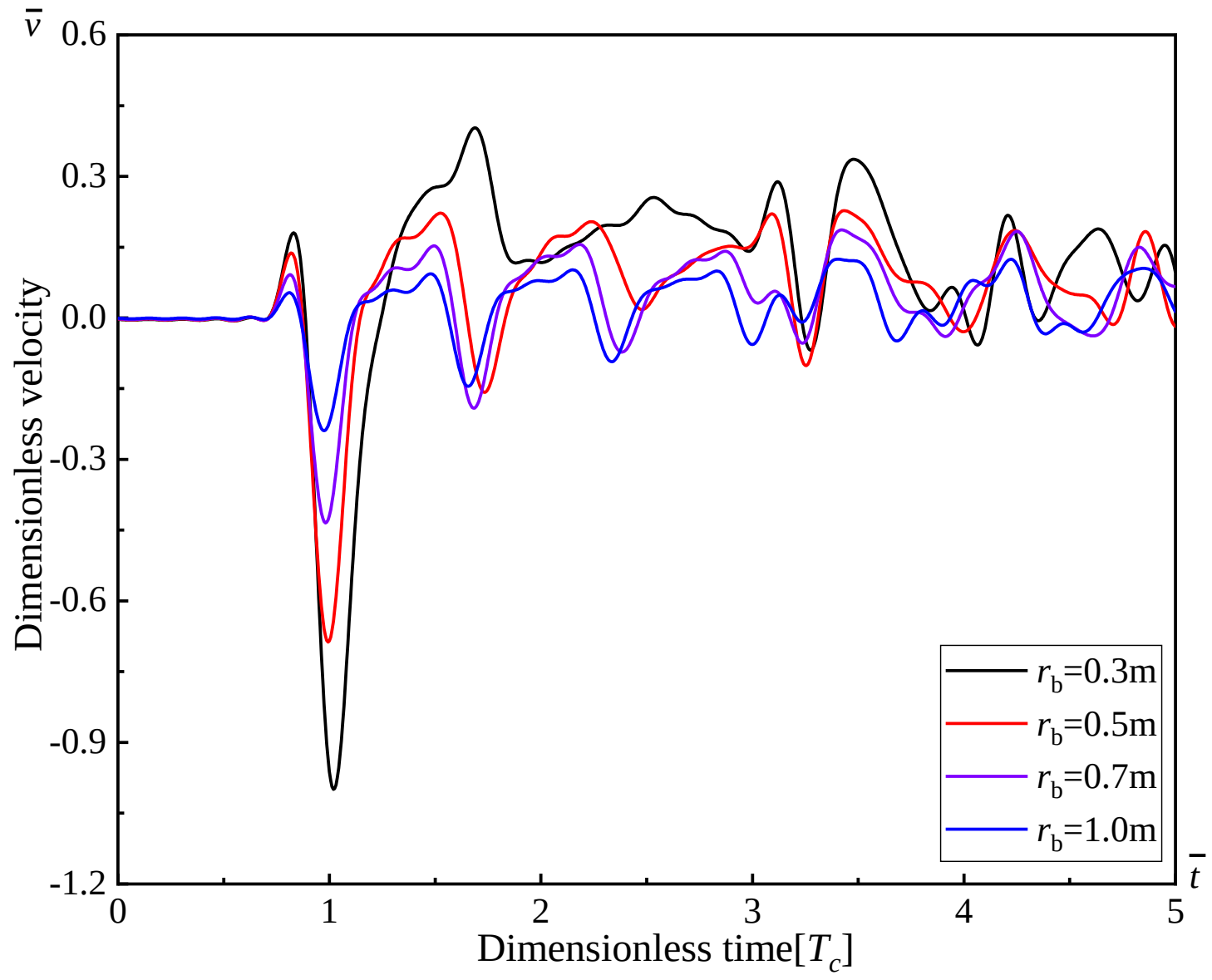

Fig.14. Effect of the sheet radius on the velocity in time domain

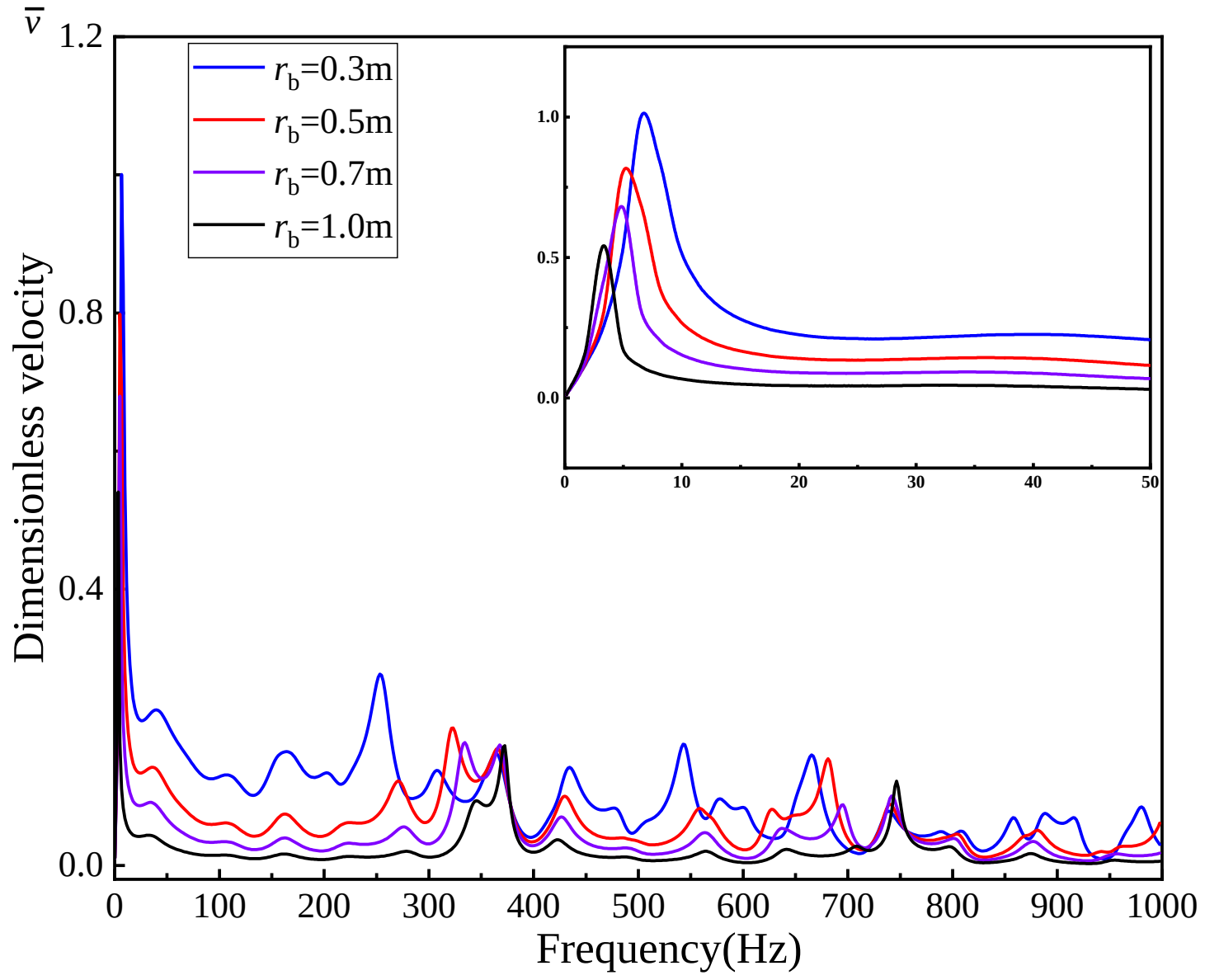

Fig.15. Effect of the sheet radius on the velocity in frequency domain

### 5.4 Length of the sheet

Fig.16 and Fig.17 demonstrate the effect of the sheet length on dynamic response in time and frequency domain, respectively. The incident wave reaches the first crest becomes earlier as the sheet length decreases. Because the sheet length gets shorter and the distance which the elastic wave travels along is reduced, resulting in a shorter arrival time to reach the first wave crest. Moreover, the velocity amplitude becomes slightly larger as the sheet length increases. Since the sheet is assumed to be a one-dimensional rod, the sheet length has little effect on the constraint effect of the joint, resulting in little difference in the velocity amplitude of the time domain.

It can be found that the amplitude-frequency curve has a high similarity in the high frequency range in Fig.17. The greater sheet length corresponds to a smaller resonance frequency in low-frequency range. It can be illustrated that the boundary condition at the end of the sheet is approximately free as the sheet length increases, which results in the reduction of the natural frequency of the sheet. Whereas, the velocity amplitude decreases as the sheet length increases. It can be understood the restrained effect of the sheet is weakened when the sheet length increases. The influence of the excitation is nearly concentrated on the front row pile, so the velocity response amplitude at the joint of the rear row pile becomes smaller.

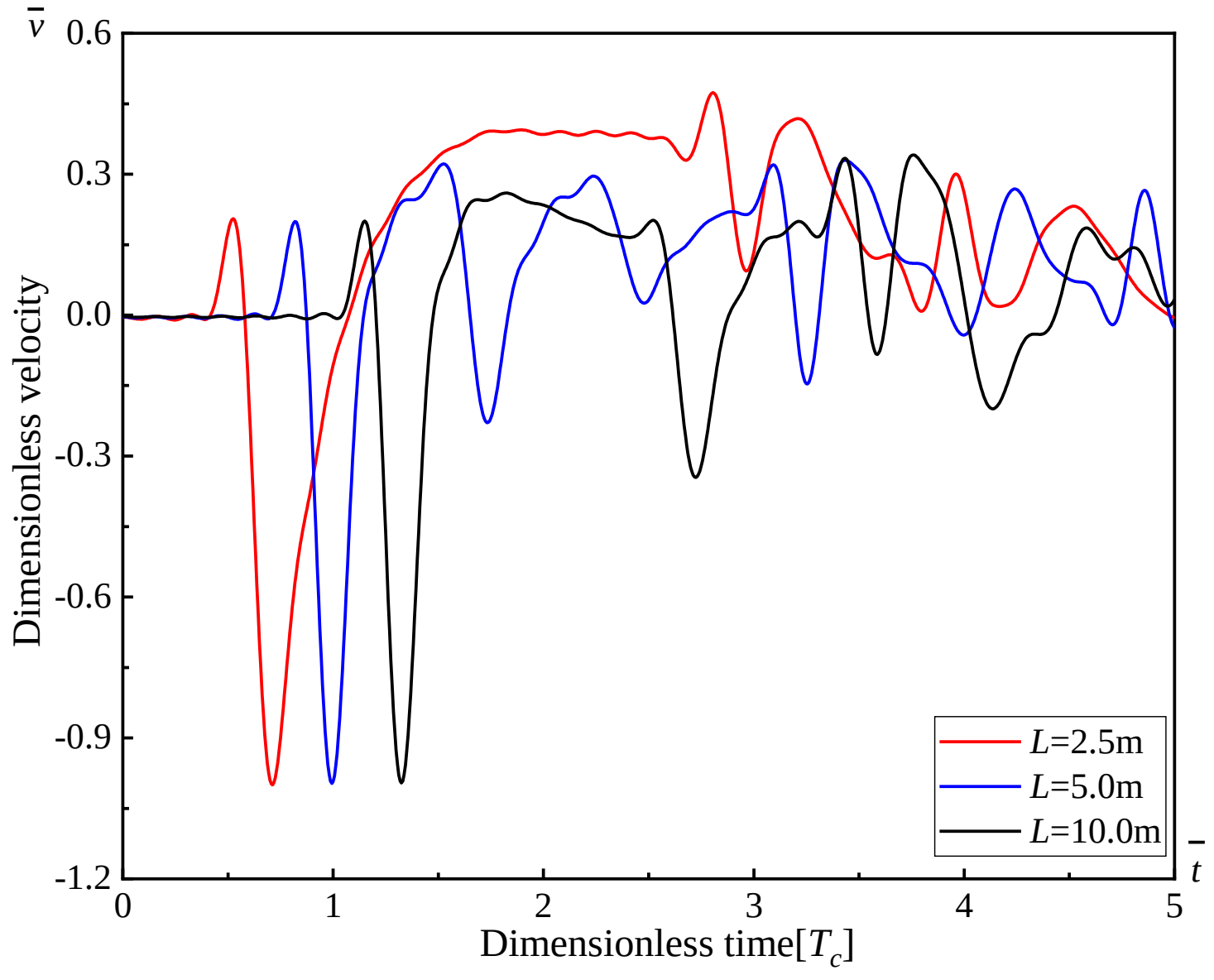

Fig.16 Effect of the sheet length on the velocity in time domain

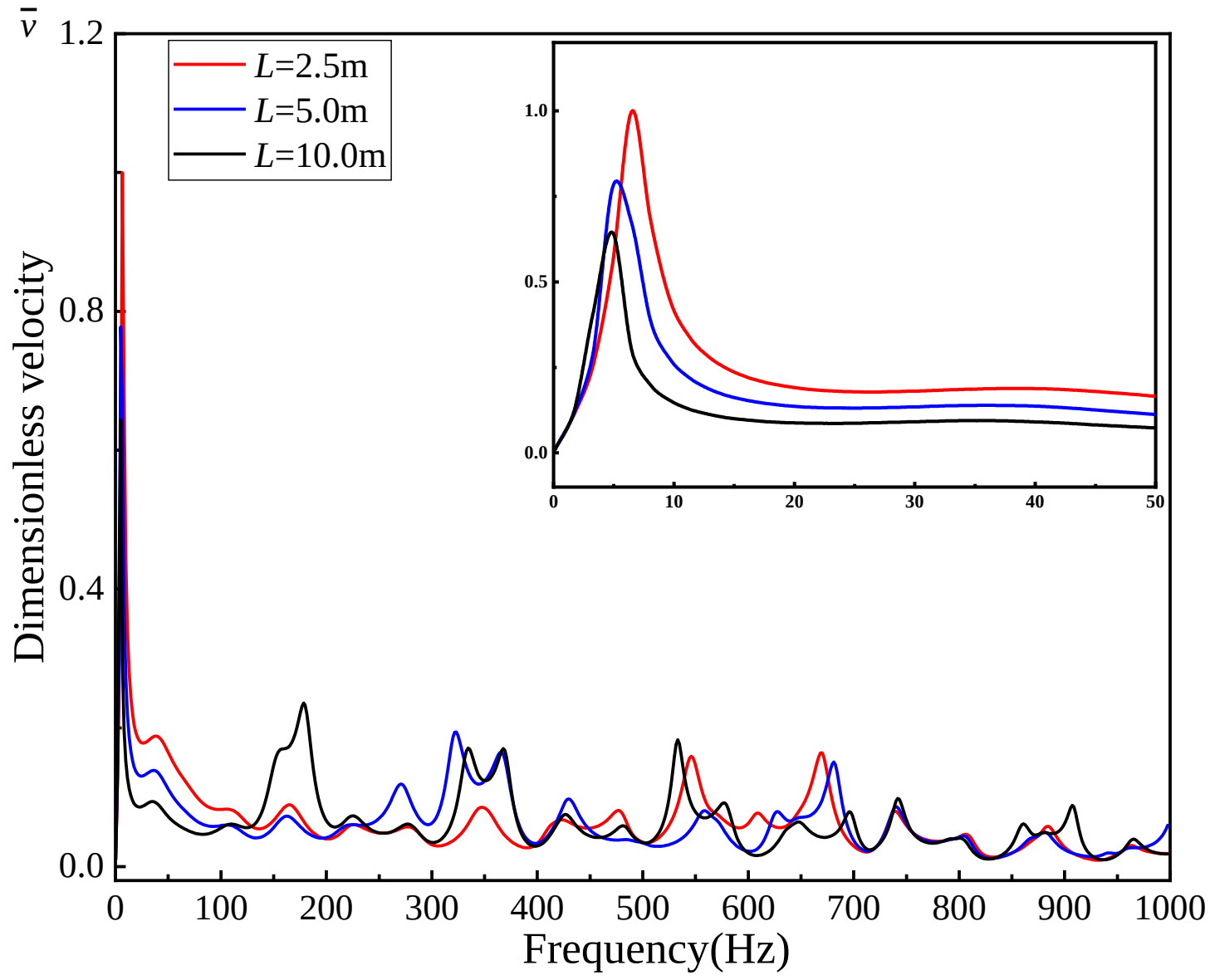

Fig.17. Effect of the sheet length on the velocity in frequency domain

### 5.5 Shear velocity of the soil

Fig.18 and Fig.19 show the effect of the shear velocity of the soil on dynamic response in time and frequency domain, respectively. The incident wave reaches the first crest almost at the same time and the velocity amplitude is almost no difference in Fig.18. It can be recognized that in the time domain, the shear wave velocity of the soil has a limited effect on the excitation on the superstructure. However, the amplitude-frequency curve has a high similarity in frequency domain in Fig.19. The greater shear velocity of the soil corresponds to a higher resonance frequency while the lower velocity amplitude corresponds to a greater shear velocity of the soil in low frequency. It can be illustrated that the greater shear wave velocity of the soil corresponds to the greater soil properties and therefore the restraint effect on the substructure becomes stronger. The sheet-pile structure is less prone to produce resonance and displacement when subjected to excitation.

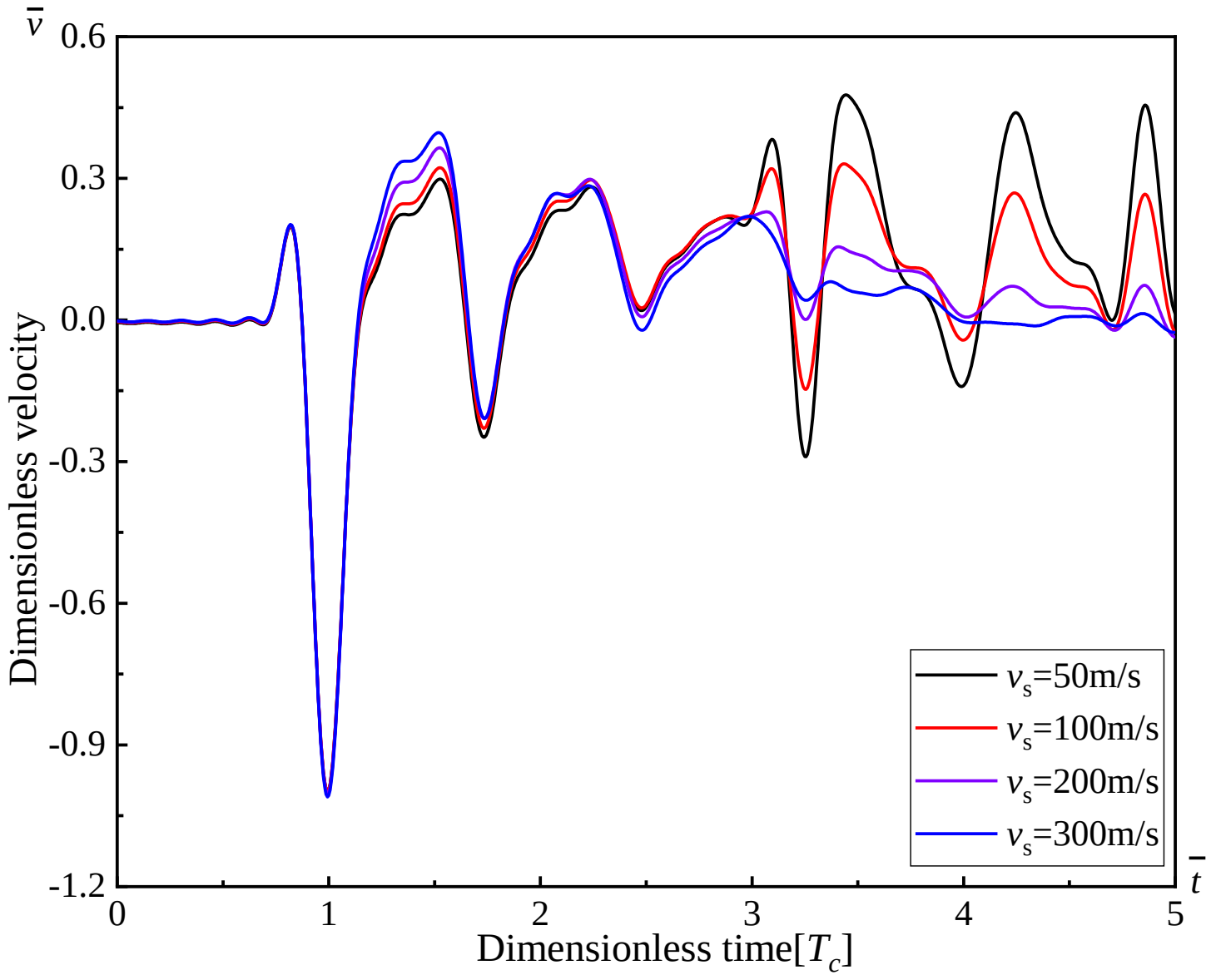

Fig.18. Effect of the shear velocity of the soil on the velocity in time domain

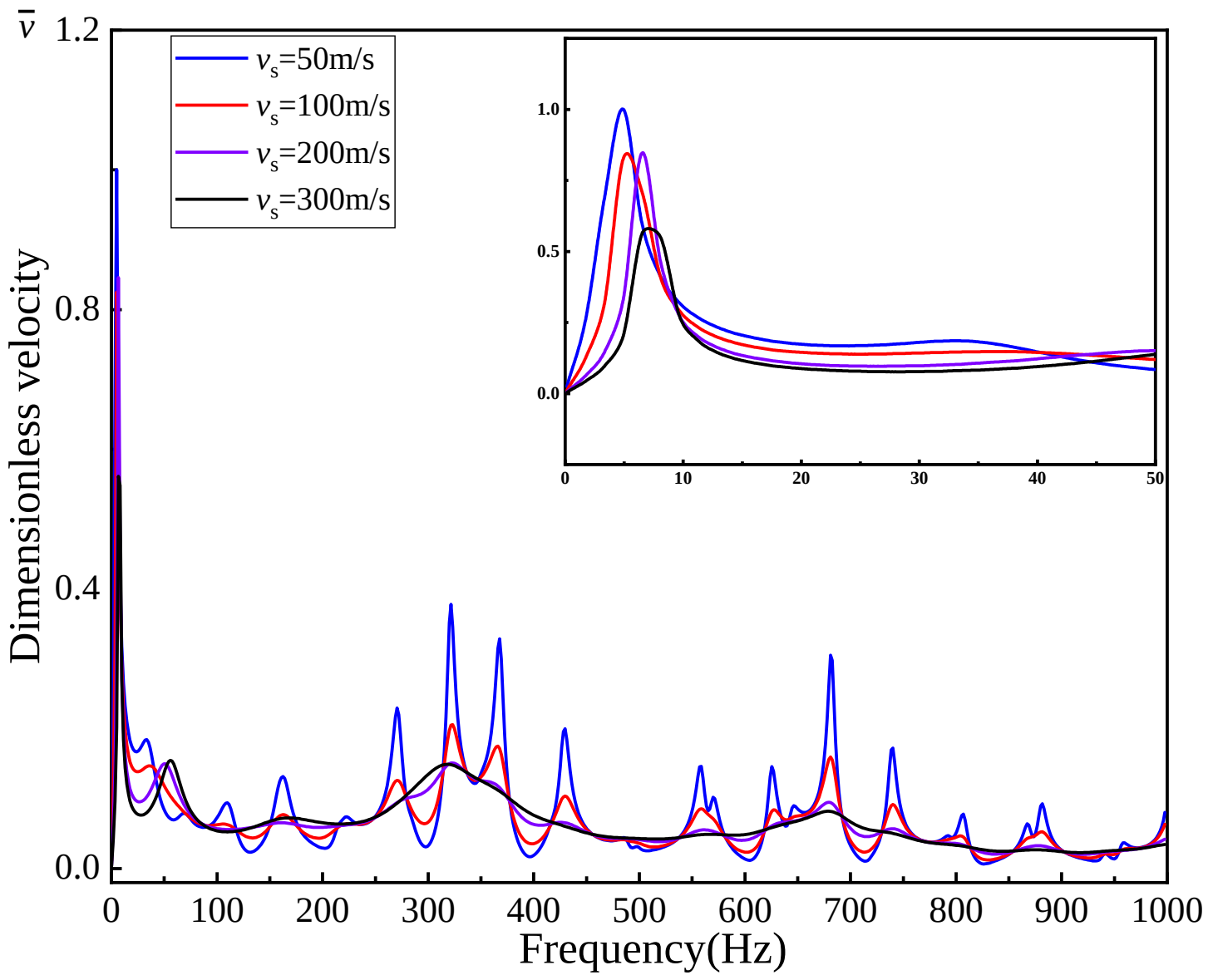

Fig.19. Effect of the shear velocity of the soil on the velocity in frequency domain

## 6. Conclusions

(1) A simplified mathematical model for the dynamic responses of the sheet-pile groin is established. The shear deformation of the pile is considered through employing TB model which also takes the internal damping into account, and the surrounding soil is idealized to the dynamic Winkler model. The sheet is simulated as a rod of which only the compression is considered.

(2) The developed model and its semi-analytical solution are verified using finite element analysis. The excellent agreement between the results obtained from both two methods verifies the accuracy of the semi-analytical solution and the reliability of the developed model.

(3) The effects of the pile radius, embedded ratio of the pile, the sheet radius, the sheet length and the shear velocity of the soil on the sheet-pile groin dynamic response in time and frequency domain are investigated through this model. The larger stiffness of the pile can explain the phenomenon that the higher resonance frequency corresponds to the greater pile radius and the smaller embedding ratio in frequency domain. Whereas, the smaller velocity amplitude corresponds to the larger stiffness of the pile in time and frequency domain. The resonance frequency becomes smaller since the constraint condition of the sheet influences the natural frequency as the radius and the length of the sheet increase. The velocity amplitude in time and frequency domain becomes smaller because the compressive rigidity of the sheet enhances the restraining effect on the movements. The increase of shear wave velocity of the soil has little effect on the velocity of the sheet-pile joints in the time domain, but it will increase the resonance frequency of the sheet-pile structure and reduce the speed amplitude in the frequency domain.

(4) Based on the research results of this article, the author suggests that designers should consider the above factors in engineering practice, i.e., avoid the resonance frequency including the corresponding maximum velocity amplitude and improve the standard to guide the safety of the structure design.

## Acknowledgments

This project has received funding from the European Union's Horizon 2020 MARIE SKŁODOWSKA-CURIE RESEARCH AND INNOVATION STAFF EXCHANGE Programme under grant agreement No 778360 and the National Natural Science Fundation of China Joint Project under grant agreement U2006225

## Appendix Ⅰ

$$\begin{cases} \eta_1 = -W_{\mathrm{p}} J_{\mathrm{p}} \\ \eta_{2n} = W_{\mathrm{p}} K_{\mathrm{p}n} - T_{\mathrm{p}} J_{\mathrm{p}} \\ \eta_{3n} = K_{\mathrm{p}n} \left( T_{\mathrm{p}} - J_{\mathrm{p}} \right) \end{cases} \tag{28}$$

$$\begin{cases} \gamma_n = \sqrt{-\dfrac{\eta_{2n}}{4\eta_1} + \sqrt{\dfrac{\eta_{3n}}{4\eta_1}}} \\ \lambda_n = \sqrt{\dfrac{\eta_{2n}}{4\eta_1} + \sqrt{\dfrac{\eta_{3n}}{4\eta_1}}} \end{cases} \tag{29}$$

$$X_{1n} = \frac{1}{T_{\mathrm{p}} - J_{\mathrm{p}}} \left[ \frac{\eta_1}{J_{\mathrm{p}}} \left( \gamma_n^2 - 3\lambda_n^2 \right) + \frac{W_{\mathrm{p}} K_{\mathrm{p}n}}{J_{\mathrm{p}}} - J_{\mathrm{p}} \right] \tag{30.a}$$

$$X_{2n} = \frac{1}{T_{\mathrm{p}} - J_{\mathrm{p}}} \left[ \frac{\eta_1}{J_{\mathrm{p}}} \left( 3\gamma_n^2 - \lambda_n^2 \right) + \frac{W_{\mathrm{p}} K_{\mathrm{p}n}}{J_{\mathrm{p}}} - J_{\mathrm{p}} \right] \tag{30.b}$$

$$T_{11}^{mn} = \mathrm{e}^{\gamma_n z} \cos\left( \lambda_n z \right), \quad T_{12}^{mn} = \mathrm{e}^{\gamma_n z} \sin\left( \lambda_n z \right) \tag{31.a}$$

$$T_{13}^{mn} = \mathrm{e}^{-\gamma_n z}\cos(\lambda_n z),\ \ T_{14}^{mn} = \mathrm{e}^{-\gamma_n z}\sin(\lambda_n z) \tag{31.b}$$

$$T_{21}^{mn} = \left[X_{1n}\gamma_n\cos(\lambda_n z) - X_{2n}\lambda_n\sin(\lambda_n z)\right]\mathrm{e}^{\gamma_n z} \tag{31.c}$$

$$T_{22}^{n} = \left[X_{1n}\gamma_n\sin(\lambda_n z) + X_{2n}\lambda_n\cos(\lambda_n z)\right]\mathrm{e}^{\gamma_n z} \tag{31.d}$$

$$T_{23}^{mn} = -\left[X_{1n}\gamma_n\cos(\lambda_n z) + X_{2n}\lambda_n\sin(\lambda_n z)\right]\mathrm{e}^{-\gamma_n z} \tag{31.e}$$

$$T_{24}^{mn} = -\left[X_{1n}\gamma_n\sin(\lambda_n z) - X_{2n}\lambda_n\cos(\lambda_n z)\right]\mathrm{e}^{-\gamma_n z} \tag{31.f}$$

$$T_{31}^{mn} = \kappa G_{\mathrm{p}}A_{\mathrm{p}}\mathrm{e}^{\gamma_n z}\left\{\left[\gamma_n\cos(\lambda_n z) - \lambda_n\sin(\lambda_n z)\right] - \left[X_{1n}\gamma_n\cos(\lambda_n z) - X_{2n}\lambda_n\sin(\lambda_n z)\right]\right\} \tag{31.g}$$

$$T_{32}^{mn} = \kappa G_{\mathrm{p}}A_{\mathrm{p}}\mathrm{e}^{\gamma_n z}\left\{\left[\gamma_n\sin(\lambda_n z) + \lambda_n\cos(\lambda_n z)\right] - \left[X_{1n}\gamma_n\sin(\lambda_n z) + X_{2n}\lambda_n\cos(\lambda_n z)\right]\right\} \tag{31.h}$$

$$T_{33}^{mn} = \kappa G_{\mathrm{p}}A_{\mathrm{p}}\mathrm{e}^{-\gamma_n z}\left\{-\left[\gamma_n\cos(\lambda_n z) + \lambda_n\sin(\lambda_n z)\right] + \left[X_{1n}\gamma_n\cos(\lambda_n z) + X_{2n}\lambda_n\sin(\lambda_n z)\right]\right\} \tag{31.i}$$

$$T_{34}^{mn} = \kappa G_{\mathrm{p}}A_{\mathrm{p}}\mathrm{e}^{-\gamma_n z}\left\{\left[-\gamma_n\sin(\lambda_n z) + \lambda_n\cos(\lambda_n z)\right] + \left[X_{1n}\gamma_n\sin(\lambda_n z) - X_{2n}\lambda_n\cos(\lambda_n z)\right]\right\} \tag{31.j}$$

$$T_{41}^{mn} = E_{\mathrm{p}}I_{\mathrm{P}}\mathrm{e}^{\gamma_n z}\left[\left(X_{1n}\gamma_n^2 - X_{2n}\lambda_n^2\right)\cos(\lambda_n z) - \left(X_{1n} + X_{2n}\right)\gamma_n\lambda_n\sin(\lambda_n z)\right] \tag{31.k}$$

$$T_{42}^{mn} = E_{\mathrm{p}}I_{\mathrm{P}}\mathrm{e}^{\gamma_n z}\left[\left(X_{1n}\gamma_n^2 - X_{2n}\lambda_n^2\right)\sin(\lambda_n z) + \left(X_{1n} + X_{2n}\right)\gamma_n\lambda_n\cos(\lambda_n z)\right] \tag{31.l}$$

$$T_{43}^{mn} = -E_{\mathrm{p}}I_{\mathrm{P}}\mathrm{e}^{-\gamma_n z}\left[\left(X_{2n}\lambda_n^2 - X_{1n}\gamma_n^2\right)\cos(\lambda_n z) - \left(X_{1n} + X_{2n}\right)\gamma_n\lambda_n\sin(\lambda_n z)\right] \tag{31.m}$$

$$T_{44}^{mn} = -E_{\mathrm{p}}I_{\mathrm{P}}\mathrm{e}^{-\gamma_n z}\left[\left(X_{2n}\lambda_n^2 - X_{1n}\gamma_n^2\right)\sin(\lambda_n z) + \left(X_{1n} + X_{2n}\right)\gamma_n\lambda_n\cos(\lambda_n z)\right] \tag{31.n}$$

## Appendix Ⅱ

$$\lambda_{\mathrm{b}} = \sqrt{\frac{\rho_{\mathrm{b}}\omega^2}{E_b + \xi_{\mathrm{b}}\omega}} \tag{32}$$

$$G_{11} = \cos(\lambda_{\mathrm{b}}x) \tag{33.a}$$

$$G_{12} = \sin(\lambda_{\mathrm{b}}x) \tag{33.b}$$

$$G_{21} = -\left(E_{\mathrm{b}}A_{\mathrm{b}} + \delta_{\mathrm{b}}\omega\right)\lambda_{\mathrm{b}}G_{12} \tag{33.c}$$

$$G_{22} = \left(E_{\mathrm{b}}A_{\mathrm{b}} + \delta_{\mathrm{b}}\omega\right)\lambda_{\mathrm{b}}G_{11} \tag{33.d}$$